\documentclass[AMA,STIX1COL]{WileyNJD-v2}

\usepackage{moreverb}
\usepackage{color, soul} 
\usepackage{graphicx}
\usepackage{url}  
\usepackage{amsthm,amsmath,bm,mathtools}
\usepackage{dsfont}
\usepackage{multirow}
\usepackage{enumerate}
\usepackage{booktabs}
\usepackage{comment}
\usepackage[font=normalsize,skip=0pt]{caption}
\newcommand{\code}[1]{\texttt{#1}}
\newcommand{\R}{{\normalfont\textsf{R }}{}}
\newcommand{\lsq}{\left[}
\newcommand{\rsq}{\right]}
\newcommand{\lbc}{\left \{ }
\newcommand{\rbc}{\right \} }
\newcommand{\lp}{\left(}
\newcommand{\rp}{\right)}

\newcommand{\cond}{{\, \vert \,}}

\articletype{Article Type}%

\received{<day> <Month>, <year>}
\revised{<day> <Month>, <year>}
\accepted{<day> <Month>, <year>}

\begin{document}

\title{Estimating heterogeneous survival treatment effect in observational data using machine learning}

\author[1]{Liangyuan Hu*}

\author[1]{Jiayi Ji}

\author[2,3]{Fan Li}

\authormark{Hu \textsc{et al}}

\address[1]{\orgdiv{Department of Population Health Science and Policy}, \orgname{Icahn School of Medicine}, \orgaddress{\state{New York}, \country{USA}}}

\address[2]{\orgdiv{Department of Biostatistics}, \orgname{Yale University School of Public Health}, \orgaddress{\state{Connecticut}, \country{USA}}}

\address[3]{\orgdiv{Center for Methods in Implementation and Prevention Science}, \orgname{Yale University School of Public Health}, \orgaddress{\state{Connecticut}, \country{USA}}}

\corres{*Liangyuan Hu, PhD\\
One Gustave L. Levy Place, Box 1077, New York, NY 10029.\\ \email{liangyuan.hu@mssm.edu}}


\abstract[Abstract]{
    Methods for estimating heterogeneous treatment effect in observational data have largely focused on continuous or binary outcomes, and have been relatively less vetted with survival outcomes. Using flexible machine learning  methods in the counterfactual framework is a promising approach to address challenges due to complex individual characteristics, to which treatments need to be tailored. To evaluate the operating characteristics of recent survival machine learning methods for the estimation of treatment effect heterogeneity and inform better practice, we carry out a comprehensive simulation study presenting a wide range of settings describing confounded heterogeneous survival treatment effects  and varying degrees of covariate overlap. Our results suggest that the nonparametric Bayesian Additive Regression Trees within the framework of accelerated failure time model (AFT-BART-NP) consistently yields the best performance, in terms of bias, precision and expected regret. Moreover, the credible interval estimators from AFT-BART-NP provide close to nominal frequentist coverage for the individual survival treatment effect when the covariate overlap is at least moderate. 
    Including a non-parametrically estimated propensity score as an additional fixed covariate in the AFT-BART-NP model formulation can further improve its efficiency and frequentist coverage. Finally, we demonstrate the application of flexible causal machine learning estimators through a comprehensive case study examining the heterogeneous survival effects of two radiotherapy approaches for localized high-risk prostate cancer.}


\jnlcitation{\cname{%
\author{Hu L}, 
\author{Ji J}, and
\author{Li F}} (\cyear{2020}), 
\ctitle{Estimating heterogeneous survival treatment effect in observational data using machine learning}, \cjournal{Statistics in Medicine}, \cvol{2020;00:1--23}.}

\maketitle


\section{Introduction}\label{sec1}

Observational data sources are expanding to include large registries and electronic health records, providing new opportunities to obtain comparative effectiveness evidence among broader target patient populations. Although traditional methods for observational studies focus on estimating the average treatment effect in a fixed target population, the rise of personalized healthcare and the increasing complexity of observational data have generated a more patient-centric view for drawing causal inferences about treatment effect.  As a result, treatments need to be tailored to the covariate distribution of the target population, which necessitates the evaluation of \emph{treatment effect heterogeneity} (TEH). If TEH presents, there exists a partition of the population into subgroups across which the treatment effect differs. Because the assessment of TEH requires addressing challenges due to complex and oftentimes high-dimensional individual characteristics, using flexible modeling techniques such as machine learning methods within the counterfactual framework is a promising approach and has received increasing attention in the statistical literature. For example, Foster et al. \cite{foster2011subgroup} studied the virtual twins approach that first uses random forests to estimate the individual treatment effect (ITE). They obtained the difference between the predicted counterfactual outcomes under different treatment condition for each unit, and then regressed the predicted ITEs on covariates to identify subgroups with a treatment effect that is considerably different than the average treatment effect. Wager and Athey \cite{wager2018estimation} developed the causal forests for ITE estimation. Lu et al. \cite{lu2018estimating} compared several random forests methods, including Breiman forests, causal forests and synthetic forests, \cite{ishwaran2014synthetic} for predicting ITE and found that the synthetic RF had the best performance. Hill \cite{hill2011bayesian} adapted the Bayesian Additive Regression Trees (BART) for causal inference and described methods for visualizing the modes of the predicted ITEs to gain insights into the underlying number of subgroups. Anoke et al. \cite{anoke2019approaches} \emph{a priori} set the number of subgroups and group observations based on percentiles of the empirical distribution of either the ITEs or propensity scores and estimated a treatment effect within each subgroup. They compared several machine learning methods including BART, gradient boosted models and the facilitating score method, for the estimation of TEH and found BART performed the best in their simulation studies. 

While these previous studies focus either on continuous or binary outcomes, time-to-event outcomes are of primary interest in many biomedical studies. \cite{hu2018modeling, ennis2018brachytherapy} Besides the average survival treatment effect, it is also important to understand whether there is survival TEH in order to make informed personalized treatment decisions. \cite{chen2018challenges} The detection of survival TEH, however, may be complicated by the right censoring (e.g. due to loss-to-follow-up) as well as challenges in precisely relating survival curves to individual covariates. Extending BART to survival outcomes, 
Henderson et al. \cite{henderson2020individualized} developed a nonparametric accelerated failure time (AFT) model to flexibly capture the relationship between covariates and the failure times for the estimation of ITE. Tian et al.\cite{tian2014simple} approached the TEH problem by modeling interactions between a treatment and a large number of covariates. Both methods were designed for analyzing data from randomized clinical trials, with little empirical evidence within the context of observational studies. Shen et al. \cite{shen2018estimating} used random forests with weighted bootstrap to estimate the optimal personalized treatment strategy; however, identification of TEH was not the focus of their paper. Cui et al.\cite{cui2020estimating} extended the causal forests of Wager and Athey \cite{wager2018estimation} for ITE estimation to accommodate right censoring, and used bootstrap for inference. Tabib and Larocque \cite{tabib2020non} also proposed a new random forests splitting rule to partition the survival data with the goal of maximizing the heterogeneity of the ITE estimates at the left and right nodes from a split. Neither of the last two methods has made software publicly available. 

Despite burgeoning literature in causal machine learning methods for survival data, their comparative performance for estimating the survival TEH remains under-explored. The practical implementation of machine learning methods for detecting possible survival TEH also merits additional study. To address these gaps and facilitate better use of machine learning for estimating survival TEH, we carry out a series of simulations to investigate the operating characteristics of nine state-of-the-art machine learning techniques, including 
\begin{enumerate}
\itemsep-0.3em
\item[(1)] nonparametric BART model within the AFT framework (AFT-BART-NP);\cite{henderson2020individualized} 
\item [(2)] a variant of AFT-BART-NP where the estimated propensity score is included as a covariate (AFT-BART-NP-PS); 
\item[(3)] semiparametric BART model within the AFT framework (AFT-BART-SP); \cite{henderson2020individualized}
\item[(4)] random survival forests (RSF);\cite{ishwaran2008random}
\item[(5)] Cox regression based on deep neural network (DeepSurv); \cite{katzman2018deepsurv}
\item[(6)] deep learning with doubly robust censoring unbiased loss function (DR-DL); \cite{steingrimsson2020deep}
\item[(7)] deep learning with Buckley-James censoring unbiased loss function (DL-BJ); \cite{buckley1979linear,steingrimsson2020deep}
\item[(8)] targeted survival heterogeneous effect estimation (TSHEE); \cite{zhu2020targeted}
\item[(9)] generalized additive proportional hazards model (GAPH) \cite{hastie1990generalized}. 
\end{enumerate}
These techniques are representative of the recent machine learning advancement specifically designed for estimating individual-specific survival functions with right-censored failure time data; details of each method are reviewed in Section \ref{sec:methods}. In particular, recent work of Dorie et al. \cite{dorie2019automated} and Hahn et al. \cite{hahn2020bayesian} suggests that inclusion of an estimated propensity score in the BART model can improve the estimation of causal effects with non-censored outcomes. To investigate whether the utility of this strategy is transferable to censored survival outcomes, we considered a version of the nonparmetric AFT BART model, AFT-BART-NP-PS, where the estimated propensity score was included as a covariate in model formulation. We contributed sets of simulations representing a variety of causal inference settings, and evaluated the empirical performance of these machine learning methods, with emphasis on new insights that can be gained, relative to the average survival treatment effects. 
We also compared these machine learning methods with two commonly used parametric and semi-parametric regression methods, the AFT and Cox proportional hazards model, and elucidated the ramifications of potentially restrictive parametric model assumptions. Finally, to improve our understanding of the survival effect under two radiotherapy approaches for treating high-risk localized prostate cancer, we applied these machine learning methods to an observational dataset drawn from the national cancer registry database (NCDB). We focused on 7330 prostate cancer patients \cite{ennis2018brachytherapy} who received either external beam radiotherapy (EBRT) combined with androgen deprivation (AD), EBRT+AD, or EBRT plus brachytherapy with or without AD, EBRT+brachy$\pm$AD, with at least 7920 cGy EBRT dose,  and explored the possible survival TEH in relation to individual covariate information. 

The remainder of the paper is organized as follows. Section \ref{sec:methods} introduces causal assumptions and estimand of interest; it further provides a concise overview of each machine learning method considered. In Section \ref{sec:simulations}, we describe the data generating processes and parameter constellations in our simulation study. Section \ref{sec:results} summarizes the comparative findings from simulation studies. 
In Section \ref{sec:example}, we provide a case study where we estimated the heterogeneous survival treatment effects of two radiotherapy approaches for high-risk localized prostate cancer patients. Section \ref{sec:discussion} concludes with a discussion.

\section{Statistical Methods} \label{sec:methods}
\subsection{Notation and causal estimand}\label{sec:setup}


Consider an observational study with two treatment conditions and $n$ observations, $i=1,\ldots,n$. Let $Z_i$ be a binary treatment variable with $Z_i=1$ for treatment and $Z_i=0$ for control, and $X_i$  be a vector of confounders measured before treatment assignment. We proceeded in the counterfactual framework, and defined the causal effect for individual $i$ as a comparison of $T_i (1)$ and $T_i (0)$, representing the counterfactual survival times that would be observed under treatment and control, respectively. We similarly define $C_i(1)$ and $C_i(0)$ as the counterfactual censoring times under each treatment condition. We further write $T_i$ as the observed survival time and $C_i$ the observed censoring time. The observed outcome consists of $Y_i=\min(T_i, C_i)$ and censoring indicator $\Delta_i=I (T_i<C_i)$. Throughout, we maintained the standard assumptions for drawing causal inference with observational survival data: \cite{chen2001causal}
\begin{enumerate}
\itemsep-0.3em
\item[(A1)] Consistency: The observed failure time and censoring time $T_i=Z_i T_i (1)+(1-Z_i) T_i (0)$ and $C_i = Z_i C_i(1) + (1-Z_i) C_i(0)$;
\item[(A2)] Weak unconfoundedness: $T_i (z) \amalg Z_i  | X_i \text{ for } Z_i= z,z\in\{0,1\}$;
\item[(A3)] Positivity: the propensity score for treatment assignment $e(X_i)=P(Z_i=1|X_i)$ is bounded away from 0 and 1;
\item[(A4)] Covariate-dependent censoring: 
$T_i (z) \amalg C_i(z) | \{X_i,Z_i\}$, for $Z_i= z,z\in \{0,1\}$.
\end{enumerate}
Assumption (A1) maps the counterfactual outcomes and treatment assignment to the observed outcomes. It allows us to write $Y_i=Z_iY_i(1)+(1-Z_i)Y_i(0)$, where $Y_i(z)=\min\{T_i(z),C_i(z)\}$. Likewise, $\Delta_i=Z_i\Delta_i(1)+(1-Z_i)\Delta_i(0)$, where $\Delta_i(z)=I\{T_i(z)<C_i(z)\}$. Assumption (A2) is referred to as the ``no unmeasured confounders'' assumption. Because this assumption is generally not testable without additional data, sensitivity analysis may be required to help interpret the causal effect estimates.\cite{hogan2014bayesian} Assumption (A3) is also referred to as the overlap assumption,\cite{li2018balancing} which requires that the treatment assignment is not deterministic within each strata formed by the covariates. This assumption can be directly assessed by visualizing the distribution of estimated propensity scores. Finally, Assumption (A4) states that the counterfactual survival time is independent of the counterfactual censoring time given the pre-treatment covariates and treatment. This condition directly implies $T_i \amalg C_i | \{X_i,Z_i\}$, and is akin to the ``independent censoring'' assumption in the traditional survival analysis literature. 


For survival outcomes, the population-level survival treatment effect can be characterized by the survival quantile effect, \cite{mao2018propensity}
\begin{equation} \label{eq:survq}
\omega^q=\theta^{q}(1)-\theta^{q}(0)=\sup\lbc t: P(T(1)\leq t) \leq q \rbc  - \sup\lbc t: P(T(0)\leq t) \leq q \rbc 
\end{equation}
where $q$ is a pre-specified number between 0 and 1. The estimand \eqref{eq:survq} compares the $q$th quantile of the marginal counterfactual survival distribution under the treated and the control condition. Implicit in our notation, we assume that the treatment strategy includes a hypothetical intervention that removes censoring until the end of follow-up. It has been suggested that the survival quantile effect should be used when the proportional hazards assumption does not hold. \cite{uno2014moving,uno2015alternatives} In this paper, we focus on $q=1/2$ and $\omega^{q=1/2}$ corresponds to the causal contrast in median survival times, which is a natural summary statistics of the entire causal survival curve and of clinical relevance. \cite{mao2018propensity,hu2019causal} Extending the population-level treatment effect \eqref{eq:survq}, we define the \emph{individual survival treatment effect} (ISTE) as
\begin{equation}\label{eq:ISTE}
\omega^{q=1/2}(x)=\theta^{q=1/2}(1,x) - \theta^{q=1/2}(0,x),
\end{equation}
where 
$\theta^{q=1/2}(z,x) = \sup\lbc t: P\lp T(z)\leq t | X=x\rp \leq 1/2\rbc$. In other words, $\theta^{q=1/2}(z,x)$ represents the median survival time of the conditional counterfactual survival function $P(T(z)>t|X=x)$ and $\omega^{q=1/2}(x)$ is alternatively called the \emph{conditional survival treatment effect}. Because only the outcome, $(Y_i, \Delta_i)$, corresponding to the actual treatment assignment is observed, the ISTE as a function of the unknown counterfactuals is not directly identifiable without additional assumptions. However, under assumptions A1-A3, we have 
\begin{align}\label{eq:identify}
P\lp T(z)>t|X=x \rp=
P\lp T(z)>t|Z=z,X=x \rp =P\lp T>t|Z=z,X=x\rp,
\end{align}
where $P\lp T>t|Z=1,X=x\rp$ becomes the group-specific survival functions conditional on $X=x$ and is estimable from observed data. Specifically, the identification condition \eqref{eq:identify} allows us to estimate $\omega^{q=1/2}(x)$ by fitting two survival outcome regression models, $\mathcal{M}_0$ and $\mathcal{M}_1$, to two sets of the observed data $\{X_i,Y_i, \Delta_i \mid Z_i=0\}$  and $\{X_i,Y_i, \Delta_i
\mid Z_i=1\}$. We then predict, for each data point, two individual-specific counterfactual survival curves conditional on individual covariate profile $X_i$, $\hat{P}\lp T_i(z)>t|X_i \rp=\hat{S}_{\mathcal{M}_z}(t,X_i)$, from model $\mathcal{M}_z, z= 0,1$. The general form of the ISTE estimator is then given by
 \begin{equation}\label{eq:medsurv}
\hat{\omega}^{q=1/2}(X_i) = \hat{\theta}_{\mathcal{M}_1}^{q=1/2}(X_i) - \hat{\theta}_{\mathcal{M}_0}^{q=1/2}(X_i),
\end{equation}
where $\hat{S}_{\mathcal{M}_1} \lp \hat{\theta}_{\mathcal{M}_1}^{q=1/2}(X_i), X_i\rp =  \hat{S}_{\mathcal{M}_0} \lp \hat{\theta}_{\mathcal{M}_0}^{q=1/2}(X_i), X_i\rp = 1/2$. 

\subsection{Estimating individual-specific counterfactual survival curves}\label{sec:overview}

Because the estimation of the causal estimand \eqref{eq:ISTE} critically depends on the estimation of individual-specific counterfactual survival functions (or conditional survival functions), $\hat{S}_{\mathcal{M}_z}(t,X)$ for $z=0,1$, below we briefly review the machine learning methods for estimating $\hat{S}_{\mathcal{M}_z}(t,X)$. For simplicity, we drop the subscript ${\mathcal{M}_z}$ in the survival function, and only note here that the flexible outcome model will be fitted for each treatment group in a symmetric fashion, and the causal interpretation will be based on \eqref{eq:identify}.


\subsubsection{Accelerated failure time BART model}
The AFT-BART model relates the failure time $T$ to the covariates through $\log T = f(X)+W$, where $f(X)$ is a sum-of-trees BART model\citep{chipman2010bart}, $f(X)=\sum_{j=1}^J g(X;\mathcal{T}_j,\mathcal{M}_j)$, and $W$ is the residual term. The sum-of-trees model is fundamentally an additive model with multivariate components, which can more naturally incorporate interaction effects than generalized additive models and more easily incorporate additive effects than a single tree model. \citep{chipman2010bart} If  $W$ is assumed to follow a Gaussian distribution, the resultant model is semi-parametric, termed as AFT-BART-SP. Henderson et al.\cite{henderson2020individualized}  model the distribution of $W$ as a location-mixture of Gaussian densities by using the centered Dirichlet process (CDP), leading to a Bayesian nonparametric specification, which we refer to as AFT-BART-NP. The individual trees $\mathcal{T}_j$ and bottom nodes $\mathcal{M}_j$ are tree model parameters. A regularization prior is placed on $\{\mathcal{T}_j,\mathcal{M}_j\}$ to allow each individual tree model to contribute only a small part to the overall fit and therefore avoids over-fitting. \cite{chipman2010bart} Furthermore, prior distributions are assumed for the distributional parameters of the residual term $W$, including the normal variance components, or the location and scale parameters for the CDP assumption. The posterior distributions of AFT model parameters are computed using Markov chain Monte Carlo (MCMC) -- Metropolis-within-Gibbs sampler for updating tree parameters and then Gibbs sampler for parameters related to the residual distribution. The unobserved survival times are imputed from a truncated normal distribution $N_{(\log Y_i,\infty)}\lp f(X_i),\sigma^2\rp$ in each Gibbs iteration, where $\sigma^2$ is the variance or common scale of the residual term. The individual-specific survival curve can then be calculated by summarizing the posterior predictive distribution of
\begin{equation*}
S^{\text{AFT-BART-SP}}(t,X_i) = 1-\Phi \lp \frac{\log t-f(X_i)}{\sigma} \rp
\end{equation*} 
for AFT-BART-SP, where $\Phi(\cdot)$ is the cumulative standard normal distribution and $\sigma$ is the variance of the residual term. For AFT-BART-NP, the individual-specific survival curve is calculated by summarizing the posterior predictive distribution of
\begin{equation*}
S^{\text{AFT-BART-NP}}(t,X_i) =
1-\sum_{l=1}^L \pi_l \Phi \lp \frac{\log t-f(X_i)-\tau_l}{\sigma} \rp,
\end{equation*}
where $\pi_l$ is the mixture proportion of cluster $l$ ($l=1,\ldots,L$), $\tau_l$ is the cluster-specific location parameter, $\sigma$ is the common scale parameter. Additional technical details are available in Herderson et al. \cite{henderson2020individualized} A recent review of BART methods can also be found in Tan and Roy. \cite{tan2019bayesian}

A direct implication from the weak unconfoundedness assumption (A2) is that $T_i(z)\amalg Z_i|e(X_i)$ for $Z_i=0,1$, suggesting that controlling for the one-dimensional propensity score is sufficient to remove confounding in estimating the average causal effect.\cite{rosenbaum1983central} For the purpose of estimating ITE with non-censored outcomes, Dorie et al. \cite{dorie2019automated} and Hahn et al. \cite{hahn2020bayesian} considered including an estimated propensity score as an additional covariate in BART for improved robustness and estimation efficiency. Assuming $\hat{e}_i$ is an estimate of $e(X_i)$, we further consider a version of AFT-BART-NP that includes both $X_i$ and $\hat{e}_i$ in the model formulation. To mitigate concerns on model misspecification, we obtain the propensity score estimate, $\hat{e}_i$, non-parametrically from a Super Learner \cite{pirracchio2015improving} and treat it as a fixed covariate in the subsequent AFT-BART-NP model formulation. The individual-specific survival curve from AFT-BART-NP-PS is then obtained by summarizing the posterior predictive distribution of
\begin{equation*}
S^{\text{AFT-BART-NP-PS}}(t,X_i, \hat{e}_i) =1-\sum_{l=1}^{L^*} \pi_l \Phi \lp \frac{\log t-f(X_i, \hat{e}_i)-\tau_l}{\sigma} \rp,
\end{equation*} 
where $\pi_l$ is the mixture proportion of cluster $l$ ($l=1,\ldots,L^*$), $\tau_l$ is the cluster-specific location parameter, and  $\sigma$ is the common scale parameter. 

\subsubsection{Random survival forests}
The RSF approach extends Breiman random forests to right-censored survival data. \cite{ishwaran2008random} The key idea of the RSF algorithm is to grow a binary survival tree $\mathcal{H}$ and construct the ensemble cumulative hazard function (CHF). Each terminal node of the survival tree contains the observed survival times and the binary censoring information for individuals who fall in that terminal node $h\in\mathcal{H}$.
Let $t_{1,h} < t_{2,h}<\ldots, t_{N(h),h}$ be the $N(h)$ distinct event times. 
The CHF estimate for node $h$ is the Nelson-Aalen estimator $\hat{\Lambda}_h = \sum_{\{l:t_{l,h}\leq t\}} {d_{l,h}}/{R_{l,h}}$, where $d_{l,h}$ and $R_{l,h}$ are the number of events and size of the risk set at time $t_{l,h}$. All individuals within node $h$ then have the same CHF estimate. To compute the CHF for an individual with covariates $X_i$, one can drop $X_i$ down the tree, and obtain the Nelson-Aalen estimator for $X_i$'s terminal node, $\Lambda(t|X_i)=\hat{\Lambda}_h(t)$ if $X_i \in h$. The out-of-bag (OOB) ensemble CHF for each individual, $\lambda^{\text{OOB}}(t|X_i)$, can be calculated by dropping OOB data down a survival tree built from in-bag data, recording $i$’s terminal node and its CHF and taking the average of all of the recorded CHFs. By a conversation-of-events  principle, \cite{ishwaran2008random} the individual-specific survival curve can be computed as one minus the estimated ensemble OOB CHF summed over survival times (event time or censoring time) until $t$, conditional on $X_i$, namely 
\begin{equation*}
S^{\text{RSF}}(t,X_i)=1-\sum_{\{j:T_j\leq t\}} \hat{\Lambda}^{\text{OOB}} (T_j|X_i). 
\end{equation*}

\subsubsection{Targeted survival heterogeneous effect estimation}

TSHEE was recently proposed based on the doubly robust estimation method, the targeted maximum likelihood estimation (TMLE), for survival causal inference. \cite{zhu2020targeted} At the first step, a stylized counting process representation of the survival data is created, then the Super Learner \cite{van2007super} is applied to estimate, for each individual at each survival time point, the conditional hazard rate $h(t|X_i)$. The initial individual survival function is then derived via the probability chain rule,
$$S(t,X_i) = \prod_{j:t_j\leq t} q(t_j|X_i).$$ 
In the second step, the initial estimator is adjusted with conditional censoring probability $P(C>t|Z,X)$ and the propensity score $e(X)$ using the one-step TMLE. \cite{cai2020one} To carry out the one-step TMLE, one replaces the univariate conditional hazard function $q(t|X)$ -- a covariate in a logistic regression working (fluctuation) model for the failure event conditional hazard -- with a high dimensional vector $\overrightarrow{q}(\bullet|X) = \lp q(t_1|X), \ldots, q(t_{\max}|X)\rp$, where $\{t_1$, \ldots, $t_\text{max}\}$ corresponds to the set of ordered unique survival times. Fitting the high-dimensional logistic regression, the one-step TMLE updates in small steps locally along the working model to iteratively update the conditional survival function of failure event to form a targeted estimator $S^{\text{TSHEE}}(t,X_i)$. We refer to Zhu and Gallego \cite{zhu2020targeted} for full technical details on TSHEE.

\subsubsection{Cox regression based on deep neural network}

DeepSurv is a Cox regression tool based on deep neural network. \cite{katzman2018deepsurv} It uses a deep feed-forward neural network to predict the effects of individual covariates on their hazard rate parameterized by the weights of the network. The input to the network is each individual's baseline covariates $X_i$. The network propagates the inputs through a number of hidden layers with weights $\xi$. The weights parameter $\xi$ is optimized to minimize the average negative log partial likelihood subject to $\ell_2$ regularization,
$$l^{\text{PL}}(\xi)\coloneqq -\frac{1}{N_E} \sum_{i:\Delta_i=1} \lsq {h}_\xi(X_i) -\log \sum_{j\in R(Y_i)} \exp\lbc {h}_\xi(X_i)\rbc \rsq +\kappa ||\xi||_2^2,$$ 
where $N_E$ is the number of observed events, $Y_i$'s are the unique observed survival times, $R(Y_i)$ is the risk set at time $Y_i$, ${h}_\xi(\cdot)$ is the log hazard function, and $\kappa$ is $L_2$ regularization parameter. The hidden layers comprise fully-connected layer of nodes, followed by a dropout layer. As a fully connected layer occupies most of the parameters, neurons develop co-dependency amongst each other during the training phase, which leads to over-fitting of training data. By dropping out individual nodes, i.e., neurons with certain probability during the training phase, a reduced network is retained to prevent over-fitting. The output of the network estimates the log hazard function $\hat{h}_\xi(X)$ within the Cox model framework. To compute the individual-specific survival curve, we first calculate the Nelson-Aalen estimate of the baseline CHF using $\hat{h}_\xi(X_i)$ outputted by DeepSurv, 
$$\hat{\Lambda}_0(t) = \sum_{i:Y_i\leq t} \dfrac{\Delta_i}{\sum_{j\in R(Y_i)} \exp\lbc\hat{h}_\xi(X_i)\rbc}.$$
We then relate the survival function to the CHF and estimate the individual-specific survival curve by
$$S^{\text{DeepSurv}} (t,X_i) = \exp \lsq -\hat{\Lambda}_0(t) \exp\lbc\hat{h}_\xi(X_i)\rbc \rsq.$$ 

\subsubsection{Censoring unbiased deep learning survival model}
A recent deep learning (DL) survival modeling technique builds upon the DL algorithms for uncensored data, but develops censoring unbiased loss  functions to replace the unobserved full data loss, resulting in two DL algorithms for censored survival outcomes – one with doubly robust censoring unbiased loss (DR-CL) and one with Buckley-James censoring unbiased loss (BJ-DL). \cite{steingrimsson2020deep} The doubly robust loss function consists of two terms, one based on uncensored outcomes and one constructed to inversely weight observed outcome information with functionals of the event or censoring probability. The loss function is doubly robust in the sense that it is a consistent estimator for the expected loss if one of the models for failure time $T$ or censoring time $C$ is correctly specified but not necessary both. A related estimator BJ-DL requires only modeling the conditional survival function for $T$ for the term of censored outcomes in the loss function, which does not share the property of double robustness with the DR-CL estimator. A survival prediction model, $S(t,X)$, can be obtained by replacing the full data loss function with either the doubly robust or Buckley-James loss function and minimizing the Brier risk. \cite{steingrimsson2020deep} Specifically, $S^{\text{DR-DL}}(t,X_i)$ minimizes the doubly robust censoring unbiased loss
$$\mathcal{L}^{\text{DR}}(t) = \frac{1}{n} \sum_{i=1}^n \frac{\delta_i(t)\lp I(Y_i\geq t)-S(t,X_i)\rp^2}{\hat{G}(Y_i|X_i)} + \frac{1}{n}\sum_{i=1}^n \lbc\frac{(1-\delta_i(t)) \mathcal{K}_{L_2,t}(Y_i,X_i,S(t,X_i))}{\hat{G}(Y_i|X_i)} - \int_0^{Y_i} \frac{\mathcal{K}_{L_2,t}(Y_i,X_i,S(t,X_i))d\hat{\Lambda}_G(u|X_i)}{\hat{G}(u|X_i)}\rbc,$$ 
and $S^{\text{BJ-DL}}(t,X_i)$ minimizes the Buckley-James censoring unbiased loss
$$\mathcal{L}^{\text{BJ}}(t) = \frac{1}{n}\sum_{i=1}^n \delta_i(t)\lp I(Y_i\geq t)-S(t,X_i)\rp^2 +\lp 1-\delta_i(t)\rp \mathcal{K}_{L_2,t}\lp C_i,X_i,S(t,X_i)\rp,$$ 
where $\delta_i (t)$ is the event indicator at time $t$ for individual $i$,  $G(u|X)$ is the conditional survival function for $C$, $\mathcal{K}_{L_2,t}(u,X_i,S(t,X_i)) = E\lsq \lp I(Y_i \geq t)-S(t,X_i)\rp^2|T\geq u, X_i\rsq$ is the Brier risk, and $\hat{\Lambda}_G(u|X_i)$ is the conditional cumulative hazard fuction for $C$. We refer the reader to Steingrimsson and Morrison \cite{steingrimsson2020deep} for additional technical details.

\subsubsection {Generalized additive proportional hazards model} 
In the absence of censoring, a flexible approach for outcome modeling and prediction is the generalized additive model (GAM).\cite{hastie1990generalized} This approach has also been extended to accommodate survival data. Specifically, the generalized additive proportional hazards model (GAPH) model \citep{hastie1990generalized} represents the hazard ratio function of the Cox proportional hazards model via a sum of smooth functions of covariates, and can be used to estimate the group-specific survival functions conditional on $X=x$. In our notation, the GAPH model is written as $\lambda(t|X)=\lambda_0 (t) \exp\left\{\eta(X)\right\},$ where $\eta(X) = \sum_{j=1}^p f_j(X_j)$ is the sum of smooth functions of each confounder $X_j$. Estimation of model functions $\eta$ proceeds by maximizing the log partial likelihood coupled with the second-order smoothness penalties for each function. In other words, the objective function is
$$l^{\text{PL}}(f_1,\ldots,f_p) + \sum_{j=1}^p\lambda_j  \int \{f_j^{''}(x)\}^2dx,$$ 
where $l^{\text{PH}}$ generically denote the log partial likelihood, and $\lambda_j$ is the regularization parameter (as $\lambda_j\rightarrow\infty$, $f_j(x)$ degenerates to the identity function). Similar to the DeepSurv, we can obtain the Nelson-Aalen estimate of the baseline CHF $\hat{\Lambda}_0(t)$ and compute the individual-specific survival curve by
$$S^{\text{GAPH}} (t,X_i) = \exp \lbc -\hat{\Lambda}_0(t) \exp\lsq \hat{\eta}(X_i) \rsq \rbc.$$

\subsubsection{Parametric accelerated failure time model}
A standard parametric model for analyzing survival data is the accelerated failure time (AFT) model. The parametric AFT model \cite{kalbfleisch2011statistical} postulates an explicit functional relationship between the covariates and the failure time $T$, $\log T = \mu +X^T\beta + \sigma W$, where $
\beta$ is a vector of coefficients describing the effects of covariates $X$ on the log transformed survival time, $\mu$ and $\sigma$ are intercept and scale parameters, respectively,  $W$ is the error term, for which a variety of distributions can be assumed. In our simulations, we assume a standard extreme value distribution and a normal distribution, for which the survival time has a Weibull distribution and a log-normal distribution, respectively. The parameters can be estimated by maximizing the likelihood function for censored data, and are denoted by $\hat{\mu}, \hat{\beta}, \hat{\sigma}$. The individual-specific survival curve can then be calculated as $$S^{\text{AFT-Lognormal}}(t,X_i) = 1-\Phi\lp\frac{\log t-\hat{\mu} -X_i^T\hat{\beta}}{\hat{\sigma}}\rp,$$ 
and 
$$S^{\text{AFT-Weibull}}(t,X_i) = \exp \lbc -\exp\lp \frac{\log t-\hat{\mu} -X_i^T\hat{\beta}}{\hat{\sigma}} \rp \rbc.$$ A good review of parametric AFT model and the associated computational algorithms can be found in Wei. \cite{wei1992accelerated}

\subsubsection {Cox proportional hazards model} 
A Cox proportional hazards model\cite{cox1972regression} relates the hazard function and the covariates $X$ through $\lambda(t|X)=\lambda_0 (t) \exp(X^T\gamma) $, where $\gamma$ is a vector of parameters describing the multiplicative effects of pre-treatment covariates on the hazard function. The model assumes that the hazard ratio is constant across the time span (the so-called proportional hazards assumption). The parameter $\gamma$ can be estimated by maximizing the partial likelihood. Similar to the procedure described for DeepSurv, we can obtain the Nelson-Aalen estimate of the baseline CHF $\hat{\Lambda}_0(t)$ and compute the individual-specific survival curve by
$$S^{\text{CoxPH}} (t,X_i) = \exp \lbc -\hat{\Lambda}_0(t) \exp\lp X_i^T\hat{\gamma} \rp \rbc.$$
Even though the Cox model includes a nonparametric specification of the baseline hazard, the performance of this approach critically depends on the parametric specification of the linear component $X_i^T\gamma$. When the true survival outcome surface involves complicated functional forms of the covariates, both the parametric AFT model and the Cox model may be subject to bias due to inaccurate model assumptions. We use these two conventional modeling techniques as a benchmark to quantify the bias that may occur in estimating the ISTE due to model misspecification.

\section{Simulation Studies}\label{sec:simulations} 
\subsection{Simulation design} \label{sec:design}
\begin{figure}[htbp]
\centering
\includegraphics[width = 0.95\textwidth]{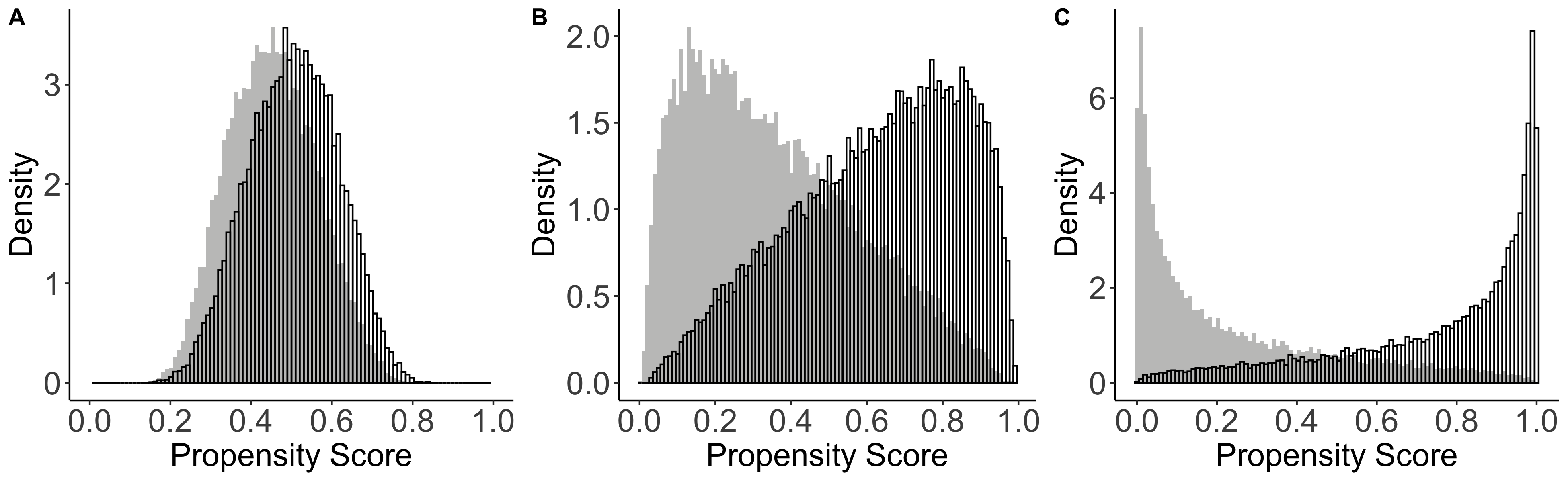}
\caption{Distributions of true propensity scores generated from three treatment assignment mechanisms with an overall treatment prevalence equal to $0.5$. The un-shaded bars indicate the treated group; the gray shaded bars indicate the control group. Panels A–C correspond to setting $\psi = 1$, $\psi = 2.5$, and $\psi=5$ respectively.}
\label{fig:overlap}
\end{figure}

We carried out a series of simulation studies to compare the above methods for estimating the individual counterfactual survival curves and hence the ISTE in the context of observational studies. Throughout we considered two levels of total sample size: $n=500$ and $n=5000$, representing small and large studies. We simulated $p=10$ independent covariates, where $X_1,\ldots,X_5$ were drawn from the standard $N(0,1)$, and $X_6,\ldots,X_{10}$ from $\text{Bernoulli}(0.5)$. The treatment assignment was generated from $Z|X \sim \text{Bernoulli}(e(X))$, where the true propensity score follows
$$\text{logit} \lbc e(X)\rbc=\alpha_0+\alpha_1 X_1+\alpha_2 X_2+\alpha_3 X_3+\alpha_4 X_5+\alpha_5 X_6+\alpha_6 X_7+\alpha_7 X_9+\alpha_8 X_{10},$$
and $(\alpha_1,\alpha_2,\alpha_3,\alpha_4,\alpha_5,\alpha_6,\alpha_7,\alpha_8 )=(-0.1\psi,-0.9\psi,-0.3\psi,-0.1\psi,-0.2\psi,-0.4\psi,0.5\psi)$. We chose $\psi\in\{1,2.5,5\}$ to represent strong, moderate and weak levels of confounding, leading to strong, moderate and weak covariate overlap. \cite{hu2020estimation,hu2021estimation} The distribution of the true propensity scores is shown in 
Figure \ref{fig:overlap}. In particular, the degree of overlap determines the effective sample size for estimating population ATE, \cite{li2019addressing} but may have important implications for estimating ISTE. The intercept in the true propensity score model $\alpha_0$ was chosen such that the proportion of treated remained approximately $0.5$. 

We simulated the true counterfactual survival times $T$ from the Weibull distribution,
\begin{equation}
P(T(z)>t|X=x)=S(t|X=x,Z=z)=\exp\lsq-\lbc d_z \exp\left(m_z(x)t\right) \rbc^\eta\rsq ~~~~\forall z\in\{0,1\},
\end{equation}
where $d_z$ is the treatment-group-specific parameter, $\eta$ is the shape parameter and $m_z(x)$ represents a generic functional form of covariates on survival time; specifications of these parameters will be discussed below. We considered the Weibull distribution because it has both the AFT and Cox proportional hazards representations, and therefore allows us to simulate scenarios where either one of these assumptions is valid. Using the inverse transform sampling, 
we generated counterfactual survival times by,  
\begin{equation*}
  T(z) = \lbc \frac{-\log U}{d_z \exp \lsq m_z(X)\rsq}  \rbc^{{1}/{\eta}}, \;\;\;\; \forall ~z \in\{0,1\},
\end{equation*}
where $U \sim \text{Unif}(0,1)$ was a random variable following a uniform distribution on the unit interval $[0,1]$, $d_z=1200$ for $z=0$, and $d_z=2000$ for $z=1$, and $\eta$ was the shape parameter used to induce non-proportional hazards. We considered three structures for $m_z(X)$, leading to four treatment effect heterogeneity settings (HS) with increasing complexities:
\begin{enumerate}[(i)]
 \item $m_1(X)=-0.2+0.1/(1+e^{-X_1})-0.8 \sin(X_3 )-0.1X_5^2-0.3X_6-0.2X_7$, and  $m_0 (X)=0.2-0.5X_1-0.8X_3-1.8X_5-0.9X_6-0.1X_7$,
 \item $m_1(X)=-0.2+0.1/(1+e^{-X_1})-0.8 \sin(X_3 )-0.1X_5^2-0.3X_6-0.2X_7$, and $m_0 (X)=-0.1+0.1X_1^2-0.2 \sin(X_3 )+0.2/(1+e^{-X_5 } )+0.2X_6-0.3X_7$,
 \item $m_1 (X)=0.5-0.1/(1+e^{-X_2} )+0.1 \sin(X_3 )-0.1X_4^2+0.2X_4-0.1X_5^2-0.3X_6$, and $m_0 (X)=-0.1+0.1X_1^2-0.2 \sin(X_3 )+0.2/(1+e^{-X_5 } )+0.2X_6-0.3X_7$,
 \item  $m_1 (X)=0.5-0.1/(1+e^{-X_2} )+0.1 \sin(X_3)-0.1X_4^2+0.2X_4-0.1X_5^2-0.3X_6$, and $m_0 (X)=-0.2+ 0.5\sin(\pi X_1 X_3 )+0.2/(1+e^{-X_5 } )+0.2X_6-0.3X_7$.
\end{enumerate}

Across all four outcome generating processes, $X_1, X_3, X_5, X_6, X_7$ were confounders that were related to both the treatment assignment and the potential survival outcome. In scenario (i), there is a nonlinear covariate effect only in the treated group. In scenario (ii), the nonlinear covariate effects are present in both treatment groups. Scenario (iii) is similar to scenario (ii) except that a non-overlapping set of covariates are included in the treated and control groups; this additional complexity represents an additional source of TEH. To avoid concerns that these settings may favor tree-based models, in scenario (iv), we further increased the data complexity by introducing a correlation of 0.6 between $X_2$ and $X_4$ in $m_1(X)$ and by including a nonadditive term $\sin(\pi X_1X_3)$ in $m_0(X)$. Finally, when simulating the counterfactual survival times, the parameter $\eta$ was set to be 2 to satisfy the proportional hazards assumption. To induce non-proportional hazards, $\eta$ was set to be $\exp\{0.7-1.8X_3+0.8X_7\}$ for $Z=0$, and  $\exp\{0.9-0.5X_1+0.5X_2\}$ for $Z=1$. For simplicity, we generated the censoring time $C$ independently from an exponential distribution with rate parameters 0.007 and 0.02, corresponding to two levels of censoring rate (CR), 20\% and 60\%. In Section \ref{sec:censoring}, we additionally investigated the robustness of findings when censoring depends on observed covariates. An exploratory sensitivity analysis of the estimated ISTE to departure from the weak unconfoundeness assumption (A2) for all methods considered was conducted in Section~\ref{sec:UMC}. The simulated counterfactual survival curves are presented in Web Figure 1.

\subsection{Implementation} 
We used the \code{abart} function from the R package \code {bart} with the default settings to implement AFT-BART-SP, and used the R package \code{AFTrees} to implement AFT-BART-NP. We used 1100 draws with the first 100 discarded as burn-in. \code{DeepSurv} was implemented in \textsf{Theano} with the Python package \code{Lasagne}. To fit RSF, we used the \code{randomForestSRC} R package with 1000 trees, \code{mtry = p/3} and a nodesize of 3. For the TSHEE method, we used the R code available at \url{https://github.com/EliotZhu/Targeted-Survival}. To implement DR-DL and BJ-DL, we used R codes provided at \url{https://github.com/jonsteingrimsson/CensoringDL}. The GAPH model was implemented using the \code{gam} function with $\code{family=cox.ph}$ from the $\R$ package $\code{mgcv}$. Super Leaner was implemented using the $\R$ package $\code{SuperLearner}$ to estimate the propensity score $\hat{e}_i$ for the method AFT-BART-NP-PS. Models in the Super Learner ensemble library included $\code{BART}$, $\code{GAM}$ and $\code{XGBoost}$.  To facilitate the application of these state-of-the-art machine learning methods to estimate ISTE, we provide example code to replicate our simulation studies in the GitHub page of the first author \url{https://github.com/liangyuanhu/ML-DL-TEH}.

\subsection{Performance metrics}

We first assessed the overall precision in the estimation of heterogeneous effects (PEHE) for each method considered using 
$$\text{PEHE}=\sqrt{\frac{1}{n} \sum_{i=1}^n \lbc \hat{\omega}^{q=1/2}(X_i) - \omega^{q=1/2}(X_i) \rbc^2 },$$ 
where $\omega^{q=1/2}(X_i)=\theta^{q=1/2}(1,X_i)-\theta^{q=1/2}(0,X_i)$ is the difference between two counterfactual median survival times calculated from each data point, and $\hat{\omega}^{q=1/2}(X_i)$ is the estimated ISTE for individual $i$. This metric was also considered in Hill \cite{hill2011bayesian} and Dorie et al. \cite{dorie2019automated}, and a smaller value of PEHE indicates better accuracy and is considered favorable. For computational considerations, we replicated each of our simulations independently $B=250$ times for $n=5000$ and $B=1000$ times for $n=500$. We summarized the mean and standard deviation (SD) of the PEHE across $B$ replications. 

To further evaluate the operating characteristics of each method in recovering treatment effect within subpopulations, we subclassified the individuals into $\mathcal{G}=\{G_1,\ldots,G_K\}$ based on $K$ quantiles of the true propensity score distribution, and calculated relative bias and root mean squared error (RMSE) for each method. This strategy was firstly considered in Lu et al. \cite{lu2018estimating} in their simulations with non-censored outcomes, and adapted in our evaluation. We choose $K=50$ to represent a large enough number of subclasses with adequate resolution. By investigating the performance of each method in distinct covariate regions with different proportions of treatment assignment and overlap, we can assess the robustness of each method to data sparsity. More concretely, given an ISTE estimator $\hat{\omega}^{q=1/2}(X)$, we define the relative bias (or percent bias) in subclass $G_k$ by
$$\text{RelBias}(k)= \frac{E\lbc \hat{\omega}^{q=1/2}(X_i)-\omega^{q=1/2}(X_i)| X_i \in G_k \rbc}{E\lbc \omega^{q=1/2}(X_i)| X_i \in G_k \rbc},~~k=1,\ldots,K.$$ 
While Lu et al.\cite{lu2018estimating} report the the bias of each estimator on its original scale, we assess the bias on the relative scale with respect to the true treatment effect (which is constant across different methods) so that the simulation results of different methods are more comparable even across different data generating processes. Under the same data generating process, however, the relative rankings of methods should be the same regardless of using the bias or relative bias, and the choice between these two metrics are immaterial.

We also define the root mean squared error (RMSE) of an ISTE estimator for subclass $G_k$ as 
$$\text{RMSE}(k) = \sqrt{E\lbc \lp \hat{\omega}^{q=1/2}(X_i)-\omega^{q=1/2}(X_i) \rp^2 \cond X_i \in G_k \rbc}, ~~~k =1, \ldots, K.$$ 

Finally, given the objective of ISTE estimation is often associated with informing individualized treatment rule (ITR), we also consider regret as a performance metric to better connect with the literature on ITR.\citep{logan2019decision} 
Specifically, the expected regret (or sometimes referred to as risk) in subclass $G_k$ is defined as 
$$\text{Regret}(k) = E \lbc \lsq  I\left( \hat{z}_{\text{opt}}(X_i) \neq z^*(X_i)\right) \times \vert {\omega}^{q=1/2} (X_i) \vert \rsq \cond X_i \in G_k\rbc, k =1, \ldots, K, $$
where $\hat{z}_{\text{opt}}(X_i) =I\{\hat{\omega}^{q=1/2}(X_i)>0 \}$ is the recommended optimal ITR based on the estimated ISTE, $z^*(X_i) =I\{{\omega}^{q=1/2}(X_i)>0\}$ is the ``true'' (infeasible) optimal ITR given $X_i$. Similar concepts were also discussed in Manski,\cite{manski2004statistical} Hirano and Porter \cite{hirano2009asymptotics} and references therein. Intuitively, the expected regret is interpreted as the average treatment benefit loss due to incorrect treatment recommendation based on the estimated ISTE, and a smaller value indicates more precise treatment recommendations.

To estimate these performance metrics, let $G_{k,b}$ contain individuals within the $q_k$ quantile of the true propensity score from replication $b$, and $N_{G_{k,b}}=\sum_{i=1}^n I(X_i\in G_{k,b})$ denote its size. We estimate the relative bias for subclass $G_k$ by 
$$\widehat{\text{RelBias}}(k) = \frac{\lbc \frac{1}{B} \sum_{b=1}^B \frac{1}{N_{G_{k,b}}} \sum_{i=1}^n I(X_i\in G_{k,b})\lp\hat{\omega}^{q=1/2}_b(X_i) -\omega_b^{q=1/2}(X_i)\rp \rbc}{\lbc \frac{1}{B} \sum_{b=1}^B \frac{1}{N_{G_{k,b}}} \sum_{i=1}^n I(X_i\in G_{k,b})\omega_b^{q=1/2}(X_i)\rbc},$$
where the true ISTE $\omega^{q=1/2}_b(X_i)$ is computed as the difference in the median survival times of two true counterfactual survival curves for each individual with covariate profile $X_i$ in the $b$th simulation iteration, and $\hat{\omega}^{q=1/2}_b(X_i)$ is the corresponding estimate based the $b$th simulated data set. The RMSE for subclass $G_k$ was estimated using 
$$\widehat{\text{RMSE}}(k) =\sqrt{\frac{1}{B}\sum_{b=1}^B \frac{1}{N_{G_{k,b}}} \sum_{i=1}^n I(X_i\in G_{k,b})\lp\hat{\omega}^{q=1/2}_b(X_i) -\omega^{q=1/2}_b(X_i)\rp ^2}.$$
Finally, the expected regret for subclass $G_k$ was estimated by
$$\widehat{\text{Regret}}(k) = \frac{1}{B}\sum_{b=1}^B \frac{1}{N_{G_{k,b}}} \sum_{i=1}^n \lbc I(X_i\in G_{k,b}) I\left( \hat{z}_{\text{opt}}(X_i) \neq z^*(X_i)\right)\vert {\omega}^{q=1/2} (X_i) \vert \rbc .$$Of note, while it is generally challenging to obtain valid confidence intervals for the ISTE using the frequentist machine learning methods directly from the model output, the Bayesian methods naturally produce coherent posterior credible intervals for the ISTE. Thereby, we summarized in Section \ref{sec:CP} the empirical coverage probability for two ``winners" of the AFT-BART methods.

\section{Simulation Results}\label{sec:results}
\subsection {Performance under strong overlap}

A total of 12 methods were considered. We first evaluated the performance of each method without the challenge due to lack of overlap. Table \ref{tab:PEHE} summarizes, for sample size  $n=5000$, the mean and standard deviation of PEHE under each configuration with strong covariate overlap (see Panel A in Figure \ref{fig:overlap}). Several patterns are immediate. First, all machine learning methods yielded substantially smaller PEHE than the parametric AFT and Cox regression models across all simulation scenarios. Second, with uncorrelated confounders and no interaction between confounders (HS(i)-(iii)), as the heterogeneity setting became more complex, machine learning methods tended to have better performance evidenced by decreasing mean and variation of the PEHE. When correlation and interaction between confounders were introduced in HS (iv), all machine learning methods but GAPH yielded similar results as for HS (ii), whereas GAPH had increased PEHE. In sharp contrast, parametric regression methods tended to perform worse, with increasing average PEHE as the heterogeneity setting became more complex. All methods show decreasing precision in estimating ISTE when the proportional hazards assumption fails to hold and under a higher censoring rate. The Cox proportional hazards model results in the largest increase in PEHE when the proportional hazards assumption is violated. In comparison, all machine learning estimators have a much smaller increase in PEHE under non-proportional hazards. This is possibly due to more complex data generating processes (the parameter $\eta$ depends on a nonlinear form of the covariates) and slightly reduced effective sample size. 
Third, among the nine machine learning approaches, AFT-BART-NP had the best overall performance across simulation configurations; AFT-BART-NP-PS had essentially the same PEHE results as AFT-BART-NP. Web Table 1 presents the PEHE for $n=500$. The findings are qualitatively similar except that the mean and variance of PEHE increase with a smaller sample size.  

\begin{table}[htbp]
\centering
\caption{Mean (and standard deviation) of PEHE for each of 12 methods and each of 8 simulation configurations for $n=5000$ with strong overlap. CR: censoring rate;  HS: heterogeneity setting; AFT-L: AFT-Lognormal; AFT-W: AFT-Weibull; ABN: AFT-BART-NP; ABNPS:AFT-BART-NP-PS;  ABS:AFT-BART-SP.} 
\begingroup
\setlength{\tabcolsep}{4.2pt}
\begin{tabular}{clcccccccccc}
\toprule
& &  \multicolumn{4}{c}{Proportional hazards} && \multicolumn{4}{c}{Non-proportional hazards}\\
\cmidrule{3-6} \cmidrule{8-11}
CR & Method & HS(i) & HS(ii) & HS(iii) &HS(iv) && HS(i) & HS(ii) & HS(iii) & HS(iv) \\
\midrule
\multirow{10}{*}{20\%} & AFT-L &	3.63 (0.14) &	4.17 (0.15) &4.13 (0.14)&4.81 (0.15) &&	3.79 (0.13)&	4.35 (0.16)	&4.41 (0.17) &4.89 (0.18)  \\
& AFT-W &	2.11 (0.12)	& 2.58 (0.12) &	2.65 (0.12)&2.94 (0.13) &&	2.31 (0.13)	& 2.84 (0.13) &	2.81 (0.13)&3.08 (0.15)\\
& CoxPH &	2.16 (0.11)&	2.61 (0.12)&	2.61 (0.13) &2.98 (0.13)&&	3.81 (0.13) &	4.38 (0.14)&	4.39 (0.13)&4.86 (0.17)\\
& ABN	& 0.51 (0.03)&	0.21 (0.02)	&0.12 (0.02) &0.23 (0.03)&&	0.72 (0.07) &	0.41 (0.04)	& 0.33 (0.04)&0.42 (0.05)\\
&ABNPS & 0.52 (0.03)& 0.22 (0.02)&	 0.13 (0.02)&0.24 (0.03) && 0.73 (0.07)&	0.42 (0.04)&0.34 (0.04)&0.43 (0.05)\\
& ABS	& 0.56 (0.04)&	0.24 (0.03)&	0.15 (0.02)	&0.26 (0.03) && 0.78 (0.08)	 & 0.44 (0.05) &	0.36 (0.04)&0.45 (0.05)\\
& RSFs &	0.53 (0.03)&	0.22 (0.02)	&0.13 (0.02)&0.24 (0.03)	&& 0.73 (0.07) &	0.42 (0.04) &	0.33 (0.04)&0.43 (0.05)\\
& DeepSurv &	0.62 (0.04) &0.29 (0.03)&	0.22 (0.03) &0.29 (0.04)&&	0.82 (0.08)	 & 0.47 (0.06) &	0.41 (0.05)&0.49 (0.06)\\
& DR-DL &	0.58 (0.04)	&0.26 (0.02)&	0.16 (0.02)&0.27 (0.04)	&&0.78 (0.07)&	0.44 (0.05)	&0.36 (0.04)&0.45 (0.05)\\
&BJ-DL	&0.57 (0.05)&	0.25 (0.02)	&0.15 (0.02)&0.26 (0.04)&&	0.77 (0.07)	&0.43 (0.05)	&0.35 (0.04)&0.44 (0.05)\\
&TSHEE&	0.60 (0.04)	&0.27 (0.03)&	0.20 (0.03)&0.26 (0.04)	&&0.80 (0.08)&	0.46 (0.06)	&0.40 (0.04)&0.48 (0.06)\\
&GAPH &0.60 (0.04)	&0.28 (0.03)	&0.20 (0.03)&0.36 (0.05)&&0.80 (0.07)	&0.48 (0.05) &0.40 (0.04)&0.58 (0.07)\\
\midrule
& &  \multicolumn{4}{c}{Proportional hazards} && \multicolumn{4}{c}{Non-proportional hazards}\\
\cmidrule{3-6} \cmidrule{8-11}
CR & Method  & HS(i) & HS(ii) & HS(iii) &HS(iv) && HS(i) & HS(ii) & HS(iii) &HS(iv)  \\
\midrule
\multirow{10}{*}{60\%}& AFT-L &	4.14 (0.14) &	4.57 (0.15) &	4.45 (0.15)&5.19 (0.20) &&	4.39 (0.18) &	4.89 (0.19) &	4.91 (0.18) & 5.88 (0.22)  \\
& AFT-W &	2.62 (0.12) &	3.22 (0.13) &	3.14 (0.13) &4.01 (0.16)&&	2.89 (0.16) &	3.71 (0.14) &	3.72 (0.13) &4.66 (0.18) \\
& CoxPH &	2.63 (0.12) &	3.21 (0.13) &	3.22 (0.13) &3.89 (0.18)&&	4.41 (0.18) &	4.83 (0.18) &	4.83 (0.18)&5.93 (0.20) \\
 & ABN	& 0.81 (0.04) &	0.31 (0.04) &	0.21 (0.02) &0.31 (0.03)&&	0.99 (0.09) &	0.51 (0.06) &	0.41 (0.04)&0.50 (0.05) \\
 &ABNPS &0.82 (0.04)& 0.32 (0.04)	&0.22 (0.02)&0.32 (0.03) &&1.00 (0.09)	&0.52 (0.06) &0.42 (0.04)&0.51 (0.05)\\
& ABS	& 0.84 (0.04) &	0.34 (0.04) &	0.24 (0.02) &0.35 (0.03)&&	1.04 (0.09) &	0.56 (0.06) &	0.45 (0.04) &0.54 (0.05)\\
& RSFs &	0.82 (0.05) &	0.32 (0.05) &	0.22 (0.02)&0.33 (0.03) &&	1.01 (0.09) &	0.52 (0.06) &	0.41 (0.04) & 0.52 (0.05)\\
& DeepSurv &	0.85 (0.04) &	0.38 (0.04) &	0.26 (0.03)&0.37 (0.04) &&	1.09 (0.11) &	0.59 (0.07) &	0.51 (0.05) & 0.62 (0.06)\\
&DR-DL	&0.84 (0.04)&0.35 (0.04)	&0.24 (0.02)&0.40 (0.03)&&	1.04 (0.09)	&0.55 (0.06)	&0.45 (0.04)&0.56 (0.05)\\
&BJ-DL	&0.83 (0.04)&	0.35 (0.04)&	0.23 (0.02)&0.38 (0.03)&&	1.03 (0.09)&	0.53 (0.06)	&0.43 (0.04)&0.54 (0.05)\\
&TSHEE&0.85 (0.05)	&0.37 (0.05)&	0.25 (0.03)&0.36 (0.04)&&	1.08 (0.10)&	0.57 (0.07)&	0.48 (0.05)&0.58 (0.06)\\
&GAPH&0.86 (0.04)&0.38 (0.04)&0.25 (0.02)&0.47 (0.05)&& 1.08 (0.09)&	0.58 (0.06)&0.48 (0.05)&0.69 (0.07)\\
\bottomrule
\end{tabular}
\endgroup
\label{tab:PEHE}
\end{table}

The relative bias, RMSE and expected regret within the $K=50$ propensity score subclasses are summarized in Web Figure 2-7. The parametric AFT and Cox models had substantially larger bias, RMSE and regret compared to the class of machine learning methods across all scenarios; and even more so when the underlying model assumptions were violated. Specifically, the Cox model had the largest relative bias, RMSE and regret under the non-proportional hazards setting and AFT-Lognormal had the largest relative bias, RMSE and regret when the true outcomes were simulated from the Weibull curves. Among the machine learning methods, AFT-BART-NP and AFT-BART-NP-PS generally maintained the lowest bias, RMSE and regret with a larger sample size ($n=5000$) and under non-proportional hazards. AFT-BART-NP-PS, while producing the similar median values of the bias, RMSE and regret as AFT-BART-NP, had less variable performance measures demonstrated by the shorter length of boxplots of each metric. This efficiency improvement due to the inclusion of a propensity score estimate echoes the findings in Dorie et al.\cite{dorie2019automated} with non-censored outcomes. 
Overall, the tightness of the range of relative bias values for machine learning methods under the most complex HS (iii) and HS (iv) when $n=5000$ demonstrate their superior performance over the traditional parametric implementations. 

\subsection{Impact of covariate overlap}
\begin{figure}[htbp]
\centering
\includegraphics[width = 0.95\textwidth]{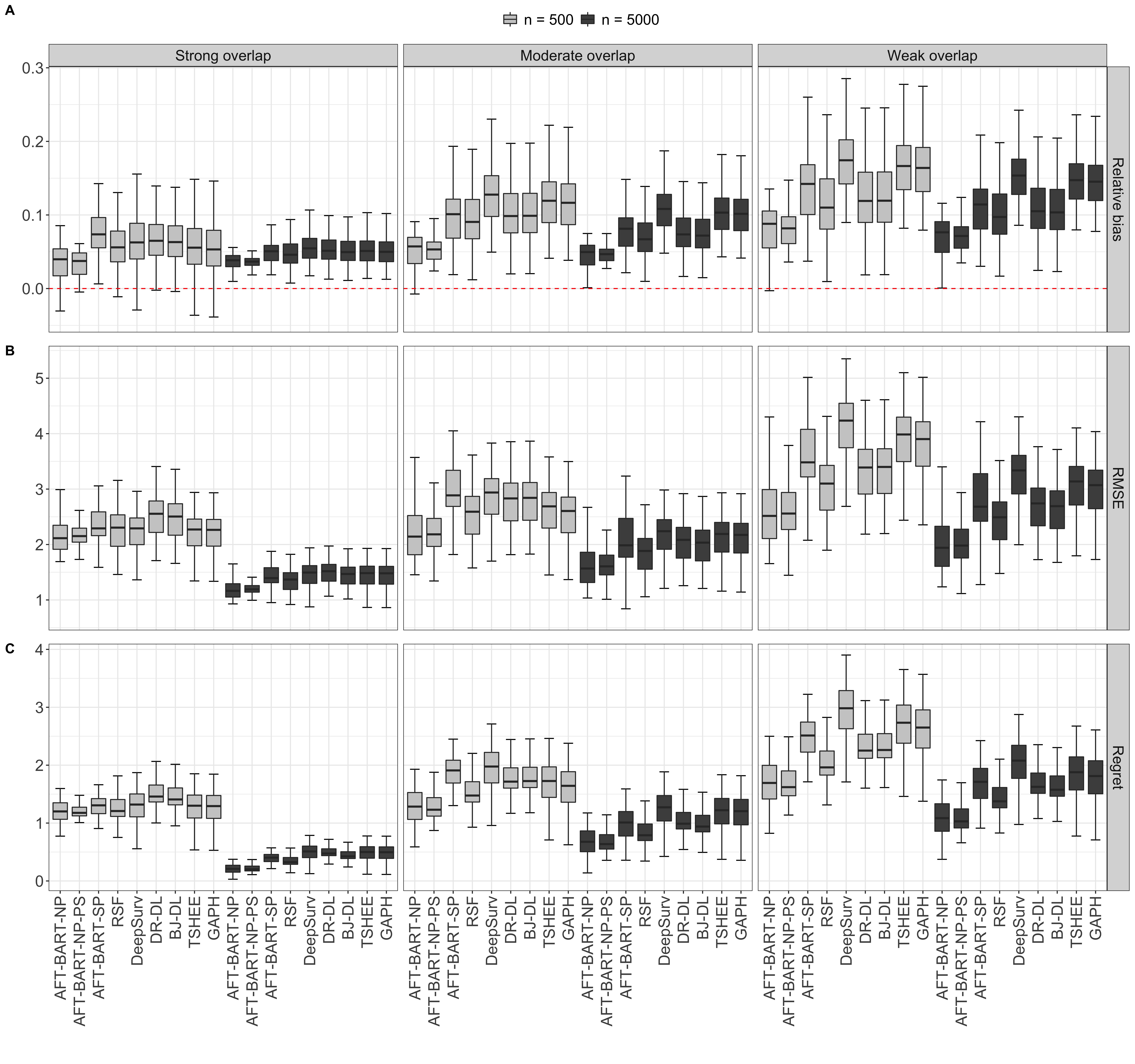}
\caption{Relative bias (Panel A), RMSE (Panel B) and expected regret (Panel C) results for each of nine machine learning methods under strong, moderate and weak covariate overlap, and for the complex scenario of heterogeneous setting (iv),  non-proportional hazards and 60\% censoring. Each boxplot visualizes the distribution of a performance measure across the propensity score subgroups $G_k$, $k=1,\ldots,50$.}
\label{fig:bias_rmse_regret_overlap}
\end{figure}

When there is strong covariate overlap, AFT-BART-NP demonstrates moderately better performance than AFT-BART-SP, RSF, DeepSurv, DR-DL, BJ-DL, TSHEE and GAPH. As the degree of covariate overlap decreases, we find that the advantage of AFT-BART-NP becomes more apparent. In addition, inclusion of the estimated propensity score in the AFT-BART-NP model considerably reduced the variability of all performance measures, regardless of degree of overlap. Figure \ref{fig:bias_rmse_regret_overlap} shows that for the most challenging scenario (HS (iv) + non-proportional hazards + 60\% censoring), the three performance measures (relative bias, RMSE and expected regret) increase as the lack of overlap increases for all machine learning methods, with both $n=500$ and $n=5000$. However, AFT-BART-NP and its variant AFT-BART-NP-PS appeared to be the most robust approach with the median relative bias around only 7\% in the most extreme scenario ($\psi=5$); DeepSurv appeared to have the largest increase in all performance metrics as the degree of overlap decreased. The two DL methods based on the censoring unbiased loss function, DR-DL and BJ-DL, showed similar performance across different simulation configurations, and performed better than the other DL method, DeepSurv, TSHEE as well as GAPH. AFT-BART-SP compared unfavorably with AFT-BART-NP, underscoring the benefit of more flexible nonparametric modeling of the residual term.   

\subsection{Overall performance assessment}
\begin{figure}[htbp]
\centering
\includegraphics[width = 0.95\textwidth]{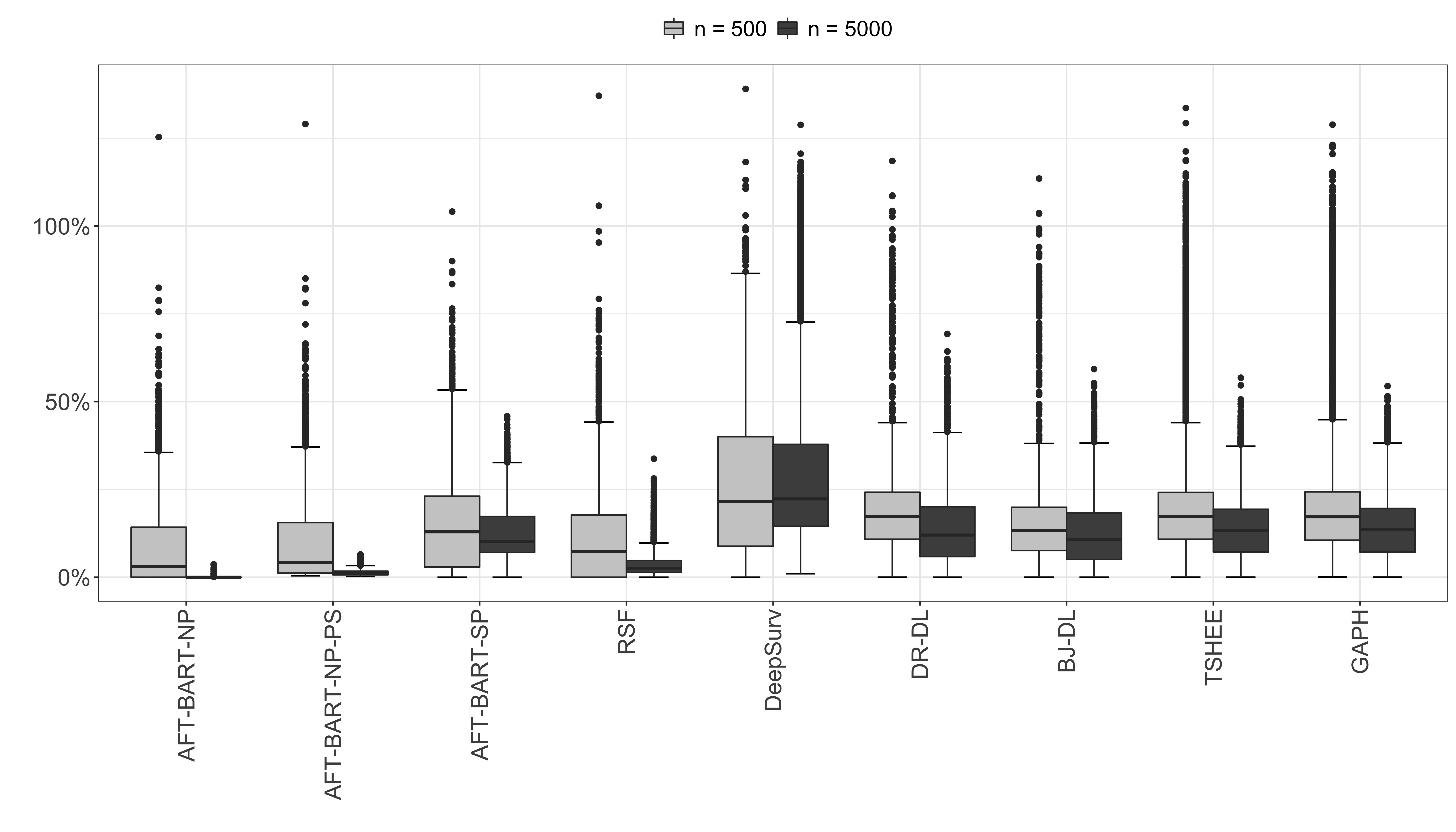}
\caption{Overall comparison of nine machine learning methods across all simulation configurations. The boxplots represent the distribution of the percent increase in PEHE with respect to the best performer for each simulation run for all runs. Smaller values are better. The total number of simulation runs for each boxplot equals to the number of scenarios multiplied by $B$ ($B=250$ for $n=5000$ and $B=1000$ for $n=500$). There are a total of 18 scenarios considered, including (a) 2 proportional hazards assumptions $\times$ 2 censoring proportions $\times$ 4 heterogeneity settings under strong overlap; plus (b) 1 proportional hazards assumptions (non-proportional hazards) $\times$ 1 censoring proportions (CR$=60\%$) $\times$ 1 heterogeneity settings (HS (iv)) under moderate and weak overlap.}
\label{fig:rmse_overall}
\end{figure}

In Figure \ref{fig:rmse_overall}, we provided an overall assessment of the machine learning methods across the simulation scenarios following the approach taken in Tabib et al. \cite{tabib2020non} Specifically, in each of the $B$ simulation iterations, we computed the percent increase in PEHE with respect to the best performer for this iteration. The percent increase in PEHE for method $s$ is given by
$$\lp\frac{\text{PEHE}_s}{\text{PEHE}_{\min}}-1\rp\times 100,$$ 
where $\text{PEHE}_{\min}$ corresponds to the PEHE from the best performer (hence with the smallest PEHE). For a given iteration, only the best method will correspond to a percent increase of exact zero while the others lead to positive values. Figure \ref{fig:rmse_overall} presents the boxplots of percent increase in PEHE over all $3500$ simulations for $n=5000$ and $14000$ simulations for $n=500$ (Details of the scenarios covered are explained in the figure title). Evidently, the AFT-BART-NP method outperformed the other approaches (with AFT-BART-NP-PS slightly trailing behind) for having the smallest median and interquartile range of the percent increase in PEHE; the advantage of AFT-BART-NP and AFT-BART-NP-PS is much more pronounced for a larger sample size $n=5000$. 

\subsection{Frequentist coverage for AFT-BART-NP}\label{sec:CP}

As we identified the AFT-BART-NP and its variant AFT-BART-NP-PS to be the ``best performer" in our simulations, we further assessed the accuracy of accodiated frequentist inference based on this flexible outcome model. An additional advantage of AFT-BART-NP is that it naturally produces coherent posterior intervals in contrast to other machine learning methods, for which it may be difficult to derive simple and valid interval estimates for ISTE. Table \ref{tab:CP} provides the mean frequentist coverage of the AFT-BART-NP credible interval estimators of the ISTE within the propensity score subclasses, under HS (iv), 60\% censoring rate and three different degrees of covariate overlap. In this most challenging HS scenario, when there is strong or moderate level of overlap, AFT-BART-NP leads to valid inference for ISTE because it generally provides close to nominal frequentist coverage across all subclasses for both sample sizes. Furthermore, under weak overlap for which it is challenging to correctly estimate or make valid inference even for the population average causal effects, \cite{li2019addressing} the AFT-BART-NP approach could still provide close to nominal frequentist coverage among the data region where treatment assignment is approximately balanced (e.g., the $25$th subclass where there is sufficient \emph{local overlap}). However, the coverage probability decreases as we move towards the tail regions of the propensity score distribution. This interesting finding on frequentist coverage echoes the previous findings for estimating the population level causal effect, but on a finer scale as we focused on individual-specific heterogeneous treatment effect. In addition, we find inclusion of the estimated propensity score in the AFT-BART model formulation further improved the mean freqentist coverage across all scenarios, and more considerably so near the tail regions of the propensity score distribution, i.e., in subclasses $G_k$'s away from $G_{25}$. A full summary of frequentist coverage across different scenarios is visualized in Web Figure 8 and conveys essentially the same message.

\begin{table}[htbp]
\centering
\caption{Summary of average frequentist coverage probability of the Bayesian credible intervals from the AFT-BART-NP (denoted by $X$) and AFT-BART-NP-PS (denoted by $X+$ PS) for 5 propensity score subclasses, for each of two sample sizes, proportional and non-proportional hazards assumptions, and under varying degrees of overlap, in the scenario of HS (iv) and 60\% censoring. The subclasses $G_k=1$ and $G_k=50$ include units with the most extreme propensity scores, with the propensity scores closest to zero in $G_k=1$ and the propensity scores closest to one in $G_k=50$. The numbers in each cell represent the mean coverage of the ISTE for the corresponding subclass.}
\begingroup
\setlength{\tabcolsep}{3pt} 
\begin{tabular}{lcccccccccccccc}
\toprule
& & \multicolumn{6}{c}{Non-proportional hazards} && \multicolumn{6}{c}{Proportional hazards}\\
\cmidrule{3-8}\cmidrule{10-15}
\multirow{3}{*}{Sample size} & \multirow{3}{*}{$G_k$} & \multicolumn{2}{c}{Strong} & \multicolumn{2}{c}{Moderate} &\multicolumn{2}{c}{Weak} &&  \multicolumn{2}{c}{Strong} & \multicolumn{2}{c}{Moderate} &\multicolumn{2}{c}{Weak} \\
& & \multicolumn{2}{c}{overlap} & \multicolumn{2}{c}{overlap} & \multicolumn{2}{c}{overlap} && \multicolumn{2}{c}{overlap} & \multicolumn{2}{c}{overlap} & \multicolumn{2}{c}{overlap} \\
\cmidrule{3-8}\cmidrule{10-15}
& & $X$  & $X +$ PS & $X$  & $X +$ PS& $X$  & $X +$ PS& &$X$  & $X +$ PS& $X$ & $X +$ PS& $X$  & $X +$ PS\\
\midrule
\multirow{5}{*}{$n=5000$}
& 1 & 0.93&0.94& 0.90&0.90& 0.50 &0.55& &0.92&0.94& 0.91&0.92& 0.49&0.56\\
 & 12 &0.92&0.94& 0.90&0.91& 0.70&0.75&& 0.92&0.93& 0.92&0.94& 0.68&0.77\\
&25& 0.94&0.95& 0.92&0.93& 0.92 &0.93&& 0.93&0.93& 0.92&0.94& 0.92&0.94\\
&37& 0.94&0.94& 0.90&0.92& 0.71 &0.75& &0.92&0.95& 0.91&0.93& 0.67&0.74\\
&50& 0.94&0.95& 0.90&0.91& 0.50 &0.56&& 0.91&0.94& 0.90&0.92& 0.50&0.57\\
\midrule
\multirow{5}{*}{$n=500$}
&1&    0.93&0.94& 0.86&0.90& 0.52&0.62&& 0.92&0.93& 0.89& 0.91& 0.52&0.58\\
& 12& 0.90&0.94& 0.90&0.90& 0.70&0.73&& 0.93&0.94& 0.90& 0.92&0.67&0.74\\
&25& 0.91&0.95& 0.91&0.92& 0.90&0.93&& 0.94&0.93& 0.92& 0.93& 0.90&0.93\\
&37& 0.90&0.94& 0.89&0.90& 0.70&0.77&& 0.92&0.94& 0.92& 0.93 &0.66&0.72\\
&50& 0.91&0.93& 0.86&0.87& 0.50&0.58&& 0.94&0.95& 0.87& 0.88& 0.50&0.57\\
\bottomrule
\end{tabular}
\endgroup
\label{tab:CP}
\end{table}

\subsection{Sensitivity to covariate-dependent censoring} \label{sec:censoring}

We assessed the sensitivity of our findings to more complex censoring mechanisms. Specifically, we replicated the simulation for the most challenging scenario (HS (iv) + non-proportional hazards) using the data generating process outlined in Section \ref{sec:design}, except that we generate the counterfactual censoring times $C(1)=C(0)$ from an exponential distribution with rate parameter $\rho(X)$, where
\begin{equation*}
    \rho(X)=0.02+0.1X_1^2+0.1\sin{(X_3)}+0.2e^{X_5}/\lp1+e^{X_5}\rp+0.02X_6-0.03X_7.
\end{equation*}
The marginal censoring proportion implied by this model remains 60\%. Web Figure 9 presents the relative bias, RMSE and expected regret across all methods under strong, moderate and weak overlap, and with sample sizes $n=500$ and $n=5000$. The results appeared almost identical to those in Figure~\ref{fig:bias_rmse_regret_overlap}, and AFT-BART-NP and AFT-BART-NP-PS were still identified to be the best approaches in terms of bias and precision. Web Table 2 presents the mean frequentist coverage of the AFT-BART-NP and AFT-BART-NP-PS credible interval estimators of the ISTE within five propensity score subclasses, in parallel to Table~\ref{tab:CP}. The results confirm that the frequentist coverage probability of these approaches is unaffected by covariate-dependent censoring.

\subsection{Sensitivity to unmeasured confounding} \label{sec:UMC}
While the flexible machine learning approaches relaxed the parametric assumptions for outcome modeling, their validity for estimating ISTE using observational data still critically depends on the structural causal assumptions (A1)-(A4) in Section \ref{sec:setup}. Among them, a key structural assumption is the availability of pre-treatment confounders so that the weak unconfoundedness holds. To assess the sensitivity of the estimated ISTE to various degrees of departure from the weak unconfoundedness assumption, we conducted an \emph{exploratory} sensitivity analysis using our simulation structure. We generated potential survival outcomes under the heterogeneous setting (iii), in which all of the machine learning methods had relatively better performance compared to other heterogeneous settings. We sequentially increased the degree of unmeasured confounding by removing access to confounders 1) $X_3$, 2) $X_3 \text{ and } X_5$ and 3) $X_3, X_5 \text{ and } X_6$ in the analysis stage for each of the 12 methods. The PEHE results for $n=5000$ are provided in Table~\ref{tab:PEHE-SA}. Results for $n=500$ convey the same message and were therefore omitted. In the presence of unmeasured confounders, the performance of all methods deteriorated, demonstrated by the larger mean and standard deviation of the PEHE under increasing degree of unmeasured confounding. However, the relative performance rankings remain the same as in the situation when the weak unconfoundedness assumption holds, with AFT-BART-NP and AFT-BART-NP-PS being the top performing methods. Results on the relative bias, RMSE and expected regret are summarized in Web Figure 10--12, and further support the findings in Table~\ref{tab:PEHE-SA}.

\begin{table}[htbp]
\centering
\caption{The mean (standard deviation) of PEHE for each of 12 methods. Data are generated under heterogeneous setting (iii) with $n=5000$ and strong overlap. $U$: unmeasured confounders; $U_1$: $X_3$ is the unmeasured confounder; $U_2$: $X_3$ and $X_5$ are the unmeasured confounders; $U_3$: $X_3$, $X_5$ and $X_6$ are the unmeasured confounders; CR: censoring rate; HS: heterogeneity setting; AFT-L: AFT-Lognormal; AFT-W: AFT-Weibull; ABN: AFT-BART-NP; ABNPS:AFT-BART-NP-PS; ABS:AFT-BART-SP.}
\begingroup
\setlength{\tabcolsep}{4.2pt} 
\begin{tabular}{clcccccccccc}\\
\toprule
& &  \multicolumn{4}{c}{Proportional hazards} && \multicolumn{4}{c}{Non-proportional hazards}\\
\cmidrule{3-6} \cmidrule{8-11}
CR &  & No $U$ & $U_1$ & $U_2$ &$U_3$ & & No $U$& $U_1$ & $U_2$ &$U_3$\\
\midrule
\multirow{10}{*}{20\%}& AFT-L & 4.13 (0.14) & 5.03 (0.16) & 5.82 (0.19) & 6.52 (0.19) && 4.41 (0.17) & 5.16 (0.20) & 5.77 (0.20) & 6.47 (0.20) \\& AFT-W & 2.65 (0.12) & 3.13 (0.14) & 3.73 (0.14) & 4.43 (0.14) && 2.81 (0.13) & 3.35 (0.16) & 4.03 (0.16) & 4.71 (0.16) \\& CoxPH & 2.61 (0.13) & 3.03 (0.14) & 3.63 (0.14) & 4.33 (0.14) && 4.39 (0.13) & 5.09 (0.20) & 5.78 (0.20) & 6.48 (0.20) \\& ABN & 0.12 (0.02) & 0.31 (0.04) & 0.46 (0.04) & 0.66 (0.04) && 0.33 (0.04) & 0.50 (0.06) & 0.65 (0.06) & 0.85 (0.06) \\ & ABNPS & 0.13 (0.02) & 0.32 (0.04) & 0.47 (0.04) & 0.67 (0.04) && 0.34 (0.04) & 0.51 (0.06) & 0.66 (0.06) & 0.86 (0.06) \\& ABS & 0.15 (0.02) & 0.35 (0.04) & 0.50 (0.04) & 0.70 (0.04) && 0.36 (0.04) & 0.53 (0.06) & 0.68 (0.06) & 0.88 (0.06) \\& RSFs & 0.13 (0.02) & 0.32 (0.04) & 0.47 (0.04) & 0.67 (0.04) && 0.33 (0.04) & 0.51 (0.06) & 0.66 (0.06) & 0.86 (0.06) \\& DeepSurv & 0.22 (0.03) & 0.38 (0.05) & 0.53 (0.05) & 0.73 (0.05) && 0.41 (0.05) & 0.58 (0.07) & 0.73 (0.07) & 0.93 (0.07) \\& DR-DL & 0.16 (0.02) & 0.35 (0.04) & 0.50 (0.04) & 0.70 (0.04) && 0.36 (0.04) & 0.53 (0.06) & 0.68 (0.06) & 0.88 (0.06) \\& BJ-DL & 0.15 (0.02) & 0.34 (0.04) & 0.49 (0.04) & 0.69 (0.04) && 0.35 (0.04) & 0.53 (0.06) & 0.68 (0.06) & 0.88 (0.06) \\& TSHEE & 0.20 (0.03) & 0.37 (0.05) & 0.52 (0.05) & 0.72 (0.05) && 0.40 (0.04) & 0.57 (0.07) & 0.72 (0.07) & 0.92 (0.07) \\& GAPH & 0.20 (0.03) & 0.37 (0.05) & 0.52 (0.05) & 0.72 (0.05) && 0.40 (0.04) & 0.57 (0.07) & 0.72 (0.07) & 0.92 (0.07) \\

\midrule
& &  \multicolumn{4}{c}{Proportional hazards} && \multicolumn{4}{c}{Non-proportional hazards}\\
\cmidrule{3-6} \cmidrule{8-11}
CR &  & No $U$ & $U_1$ & $U_2$ &$U_3$ & & No $U$& $U_1$ & $U_2$ &$U_3$\\
\midrule

\multirow{10}{*}{60\%}& AFT-L & 4.45 (0.15) & 5.33 (0.20) & 6.04 (0.22) & 6.78 (0.25) && 4.91 (0.18) & 5.99 (0.22) & 6.56 (0.22) & 7.27 (0.22) \\& AFT-W & 3.14 (0.13) & 4.22 (0.16) & 4.93 (0.19) & 5.66 (0.22) && 3.72 (0.13) & 4.75 (0.18) & 5.45 (0.18) & 6.05 (0.18) \\& Cox PH & 3.22 (0.13) & 3.94 (0.18) & 4.52 (0.20) & 5.13 (0.22) && 4.83 (0.18) & 6.02 (0.20) & 6.64 (0.20) & 7.24 (0.20) \\& ABN & 0.21 (0.02) & 0.34 (0.03) & 0.49 (0.04) & 0.69 (0.05) && 0.41 (0.04) & 0.56 (0.05) & 0.71 (0.05) & 0.92 (0.05) \\& ABNPS & 0.22 (0.02) & 0.35 (0.03) & 0.50 (0.04) & 0.70 (0.05) && 0.42 (0.04) & 0.57 (0.05) & 0.73 (0.05) & 0.93 (0.05) \\& ABS & 0.24 (0.02) & 0.38 (0.03) & 0.52 (0.04) & 0.72 (0.05) && 0.45 (0.04) & 0.59 (0.05) & 0.73 (0.05) & 0.94 (0.05) \\& RSFs & 0.22 (0.02) & 0.36 (0.03) & 0.51 (0.04) & 0.71 (0.05) && 0.41 (0.04) & 0.60 (0.05) & 0.75 (0.05) & 0.94 (0.05) \\& DeepSurv & 0.26 (0.03) & 0.40 (0.04) & 0.55 (0.05) & 0.75 (0.06) && 0.51 (0.05) & 0.68 (0.06) & 0.83 (0.06) & 1.02 (0.06) \\& DR-DL & 0.24 (0.02) & 0.44 (0.03) & 0.59 (0.04) & 0.79 (0.05) && 0.45 (0.04) & 0.65 (0.05) & 0.80 (0.05) & 1.00 (0.05) \\& BJ-DL & 0.23 (0.02) & 0.42 (0.03) & 0.56 (0.04) & 0.76 (0.05) && 0.43 (0.04) & 0.64 (0.05) & 0.79 (0.05) & 0.99 (0.05) \\& TSHEE & 0.25 (0.02) & 0.41 (0.04) & 0.56 (0.05) & 0.77 (0.06) && 0.48 (0.05) & 0.67 (0.06) & 0.82 (0.06) & 1.02 (0.06) \\& GAPH & 0.25 (0.02) & 0.41 (0.04) & 0.56 (0.05) & 0.77 (0.06) && 0.48 (0.05) & 0.67 (0.06) & 0.82 (0.06) & 1.02 (0.06) \\
\bottomrule
\end{tabular}
\endgroup
\label{tab:PEHE-SA}
\end{table}

\section{CASE STUDY: Estimating HETEROGENEOUS SURVIVAL Causal EFFECT OF TWO RADIOTHERAPY APPROACHES FOR patients with HIGH-RISK PROSTATE CANCER}\label{sec:example}

Two popular radiotherapy based approaches for treating high-risk localized prostate cancer are EBRT+AD and EBRT+brachy$\pm$AD. \cite{ennis2018brachytherapy} It is unclear whether EBRT+AD and EBRT+brachy$\pm$AD  have significantly different survival effect among those who received at least 7920 cGY EBRT dose, and whether there exists TEH. To better understand the comparative overall and heterogeneous survival causal effects of these two radiotherapy approaches, we pursued a two-stage causal analysis with the NCDB data including 7730 high-risk localized prostate cancer patients who were diagnosed between 2004 and 2013 and were treated with either EBRT+AD or EBRT+brachy$\pm$AD with no less than 7920 cGY EBRT dose. The pre-treatment confounders included in this analysis were age, prostate-specific antigen (PSA), clinical T stage, Charlson-Deyo score, biopsy Gleason score, year of diagnosis, insurance status, median income level (quartiles on the basis of zip code of residence), education (based on the proportion of residents in the patient's zip code who did not graduate high school), race, and ethnicity. Web Table 3 summarizes the baseline characteristics of patients in the two treatment groups. Clearly, there are a number of covariates that are unbalanced between the two groups, suggesting the need to adjust for confounding. Web Figure 13 shows the distribution of the estimated propensity scores by fitting a Super Learner\cite{pirracchio2015improving} on the treatment and covariates data. Relatively strong covariate overlap is suggested from the propensity score distribution.

\begin{figure}[htbp]
\centering
\includegraphics[width = 0.8\textwidth]{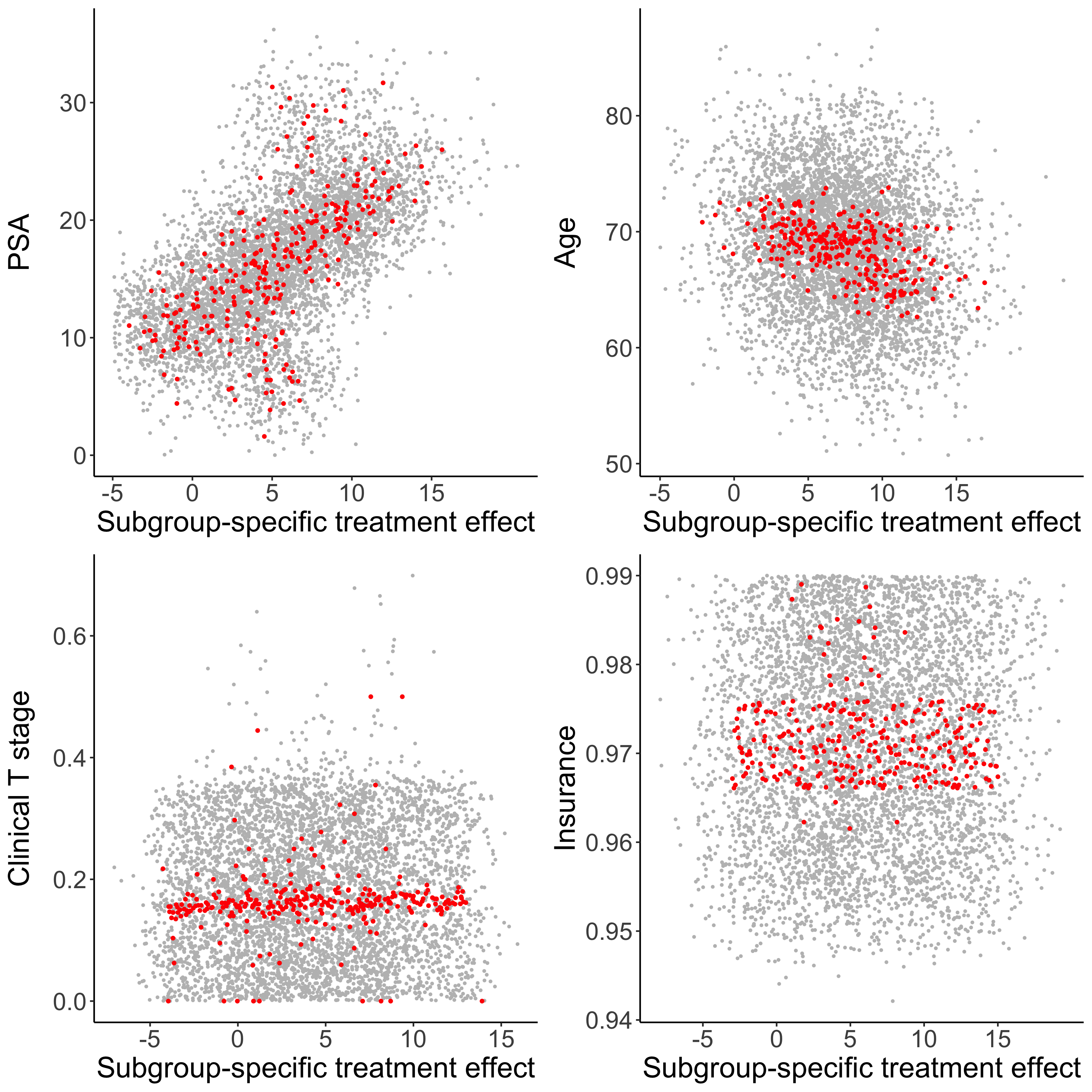}
\caption{Visualization of results from the NCDB data analysis. Covariates considered are PSA, age, clinical T stage, and insurance status. The x-axis represents the subgroup-specific posterior average survival treatment effect (median counterfactual survival in months), and the y-axis represents the subgroup-specific average value of the covariate. A red dot represents a subgroup, generated by dividing percentiles of the distribution of posterior mean individual survival treatment effects (ISTE). There are $300$ subgroups in total. Gray dots represent a random subset of $100$ (out of $1000$) posterior draws of the ISTE with the sub-grouping process applied.}
\label{fig:p_scatter}
\end{figure}

Our analyses proceed in a two-stage fashion. In the first stage, we applied the machine learning methods to estimate the ISTE within the the counterfactual framework. The estimated ISTE were then used, in the second stage, as dependent variables in an exploratory linear regression analysis to generate some further insights. As argued in Lu et al.,\cite{lu2018estimating} the estimated regression coefficients can be interpreted as how treatment effects are modulated by corresponding covariates. To estimate standard errors and uncertainty intervals for the regression coefficients, we drew for each individual $1000$ posterior MCMC samples of ISTE from the AFT-BART models. For the frequentist machine learning approaches (RSF, DeepSurv, TSHEE, DR-DL and BJ-DL), we subsampled the entire procedure to obtain uncertainty intervals for the linear regression coefficients. Specifically, we drew a sample of size $n/5$ without replacement, refit each model based on the subsampled data and used the resulting ISTE estimates as dependent variable in the exploratory linear regression. The subsampling process was repeated 1000 times. Similar to Politis et al. \cite{politis1999subsampling} and Lu et al.,\cite{lu2018estimating} we used the subsampling approach instead of bootstrap due to considerations on computational speed and robustness. Web Table 4 presents the regression coefficients and $95\%$ credible intervals for all covariates and the intercept from the AFT-BART-NP-PS model; the intercept could be considered as an overall measure of the treatment effect for the ``baseline" population (with all covariates set to be the reference category or mean value). Results from using ISTE estimated by frequentist machine learning approaches were similar and therefore omitted. The intercept estimates suggest that the population average survival treatment effect effect of EBRT+brachy$\pm$AD versus EBRT+AD is not statistically significant at the 0.05 level. However, PSA and age significantly modulated the comparative effect of the two radiotherapy approaches on median survival. For example, the estimated coefficient was 3.99 (95\% CI: $3.60 - 4.38$) for PSA; this means that difference in the counterfactual median survival under EBRT+brachy$\pm$AD versus EBRT+AD, which was positive, became greater for patients with higher PSA. Similarly, the negative coefficient estimate for age suggests that the median survival benefit from EBRT+brachy$\pm$AD were more pronounced among younger patients. In Figure \ref{fig:p_scatter}, we created $300$ subgroups based on percentiles of the distribution of posterior mean ISTE, and provide the scatter plots of subgroup-specific average covariate (PSA, age, clinical T stage and insurance status) versus subgroup-specific average survival treatment effect. While the subgroup survival treatment effect estimates are randomly scattered across values of clinical T stage and insurance status, there is a clear non-random pattern between subgroup-specific survival treatment effect estimates and PSA or age, with directions in close agreement to findings from the exploratory linear regression analysis (Web Table 4). 

\begin{figure}[htbp]
\centering
\includegraphics[width = 0.8\textwidth]{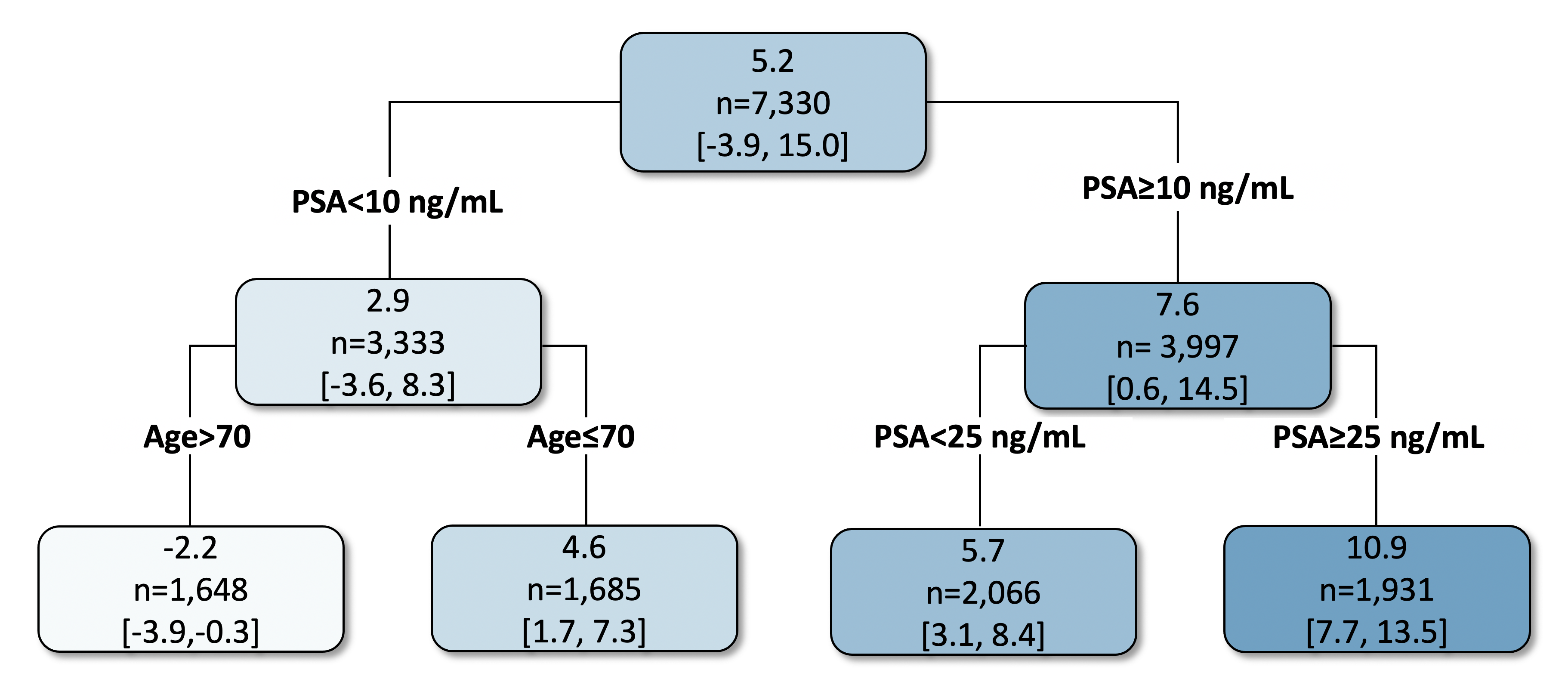}
\caption{Final CART model fit to the posterior mean of the individual survival treatment effect (measured by median survival in months) comparing EBRT plus brachytherapy with or without AD (EBRT+brachy$\pm$AD) versus EBRT plus AD (EBRT+AD). Values in each node correspond to the posterior mean and 95\% credible intervals for the average survival treatment effect for the subgroup of individuals represented in that node.}
\label{fig:fit-the-fit}
\end{figure}

To further explore the subgroups that may have enhanced survival treatment effect from either of the two radiotherapy approaches, we leveraged the posterior MCMC draws of the ISTE from AFT-BART-NP, and employed the ``fit-the-fit" strategy to explore the potential TEH in relation to the covariate subspaces. The ``fit-the-fit" strategy has been used to identify possible subgroups defined by combination rules of covariates that have differential treatment effects; see for example, Foster et al.\cite{foster2011subgroup} and Logan et al.\cite{logan2019decision} Instead of a linear regression analysis, the classification and regression tree (CART) model was used to regress the posterior mean of the ISTE on the covariates. A sequence of CART models were fit, where covariates were sequentially added to the CART model in a stepwise manner to improve the model fit, as measured by $R^2$. At each step, the variable leading to the largest $R^2$ improvement was selected into the model and the procedure ended until the percent improvement in $R^2$ was less than 1\%. This complexity parameter controls the size of the final CART tree. Subgroup treatment effects were estimated by averaging the ISTE among individuals falling into each node of the final CART model, and the branch decision rules (i.e., binary splits of the covariate space) suggested combination rules of covariates. Figure \ref{fig:fit-the-fit} shows the results of the final tree estimates. The final $R^2$ between the tree fit and the posterior mean treatment effect of EBRT+brachy$\pm$AD versus EBRT+AD was 80\%. The first splitting variable was PSA. Patients with PSA $\geq 10$ ng/ml had approximately 7.6 (95\% CI: 0.6--14.5) months longer median survival with EBRT+brachy$\pm$AD, while patients with PSA $<10$ ng/ml did not experience significantly higher median survival from either treatment. A second level of variable splitting provided further resolution on the magnitude of the treatment benefit for patients with higher PSA $\geq 10$ ng/ml; the most beneficial subgroup was those with PSA $\geq 25$ ng/ml and had approximately 10.9 months longer median survival. Among patients with lower PSA $<10$ ng/ml, older patients aged above 70 experienced treatment benefit from EBRT +AD with median survival approximately 2.2 months longer; on the contrary, younger patients aged below 70 could benefit from EBRT+brachy$\pm$AD with approximately 4.6 months longer median survival. This alternative analysis strategy generated evidence that converged to the exploratory linear regression analysis on the ISTE estimates, and suggested interpretable patient subpopulations with beneficial survival treatment effects due to EBRT+brachy$\pm$AD. We further explored the sensitivity of the summarizing CART model fit to the tuning parameters of both CART and AFT-BART-NP. When the complexity parameter value of CART ranges from 0.6\% to 1.1\%, the final tree remains the same. With a value $>1.1\%$, the final summarizing tree includes only PSA and the final $R^2$ drops to 0.67. When the complexity parameter value is $<0.6\%$, the final summarizing tree gets more complex by including an additional variable biopsy Gleason score with only a slight increase (0.005) in the final $R^2$. Varying values for the hyperparameters of the AFT-BART-NP model did not change the results of the CART model fit. Hence, the summarizing CART model fit in this case study is robust to the tuning parameters.

\section{DISCUSSION}\label{sec:discussion}

Identification of TEH has been conventionally carried out by first enumerating potential effect modifiers with subject-matter experts and then estimating the average treatment effect within each subgroup.\cite{kent2020predictive,wang2007statistics} In randomized clinical trials, this approach could also facilitate pre-specification and therefore is particularly suitable to \emph{confirmatory TEH analysis}.\cite{kent2020predictive} In observational data with complex heterogeneity of treatment effect, such \emph{a priori} specification that separates the issues of confounding and TEH is often practically infeasible, especially with a large number of pre-treatment covariates. Such complications motivate \emph{exploratory TEH analysis} for generating scientifically meaningful hypotheses and treatment effect discovery, which is the focus of this current paper.\cite{kent2020predictive,assmann2000subgroup,wang2007statistics} Exploiting on the counterfactual framework, flexibly modeling of the outcome has been shown to be well-suited for exploratory TEH analysis. However, ongoing discussions about causal methods for estimating TEH in observation data have largely focused on non-survival data. For example, tree-based methods such as BART and random forests have recently emerged as popular approaches to causal effect estimation for continuous or binary outcomes.\cite{lu2018estimating,hill2011bayesian} On the other hand, several modern deep learning methods (DeepSurv, DR-DL, BJ-DL) have been developed for censored survival data \citep{katzman2018deepsurv,steingrimsson2020deep}, and a TMLE based method has recently been proposed to estimate the survival heterogeneous effect. \citep{zhu2020targeted}
 The GAPH model is adept at 
flexibly fitting complex main effects, which are often of great interest in health studies.\citep{hastie1990generalized} We adapt these  machine learning modeling tools for estimating ISTE and provide new empirical evidence using simulations representative of complex confounding and heterogeneity settings with right-censored survival data.  

Our results indicate that all machine learning methods substantially outperformed the traditional AFT and Cox proportional hazards models in estimating TEH. This is not surprising given that the traditional models often encode strong parametric and linearity assumptions, which are prone to bias when model assumptions fail to hold. Among the class of machine learning methods, the nonparametric version of the AFT-BART model \cite{henderson2020individualized} generally outperformed the other methods, both from the traditional prediction perspective (bias and RMSE) and the decision perspective (expected regret). The advantage of AFT-BART-NP becomes more pronounced with a larger sample size and increased data complexity. On the other hand, AFT-BART-NP also conveniently enables posterior inference by providing credible intervals for ISTE and the associated subgroup-specific treatment effects. In particular, we demonstrated that when there is strong or moderate overlap, AFT-BART-NP could provide close to nominal frequentist coverage for almost all average causal effect among subpopulations defined by propensity score strata. Under weak overlap, AFT-BART-NP still provides satisfactory frequentist coverage for the subpopulation near the centroid of the propensity score distribution. While the overall population may exhibit weak overlap, the central region of the propensity score distribution may still present satisfying \emph{local overlap} and in fact admits valid inference for the causal effects. However, AFT-BART-NP could understate the uncertainty of ISTE estimates in regions with lack of overlap. This observation was also discussed in an invited Commentary to Hahn et al.\cite{hahn2020bayesian} by Papadogeorgou and Li,\cite{Papadogeorgou2020BA} who illustrated that BART for continuous outcomes produced overly narrow confidence band in the region of poor overlap. As a potential improvement, we found that including a non-parametrically estimated propensity score in the AFT-BART-NP model formulation leads to more accurate estimation and inference for ISTE, evidenced by less variable subgroup-specific performance measures, as well as higher frequentist coverage under weak overlap.
Because previous simulations for TEH rarely considered the effect of overlap, our results offer new insights. Importantly, our findings can provide new perspectives into the estimation of ISTE relative to average treatment effects. At the minimum, we noticed that even though it is challenging to estimate ISTE for units at the tails of the propensity score distribution, it remains practically feasible to make accurate inference for ISTE within the centroid of the propensity score distribution. In fact, the centroid region of the propensity score distribution includes individuals at clinical equipoise (and for whom the treatment decisions are mostly unclear) and resemble those recruited in a randomized controlled trial. In the propensity score literature, this subpopulation is sometimes referred to as the \emph{overlap population}, for which there exists an efficient weighting estimator for the average causal effect. \cite{li2018balancing,li2019addressing,li2019propensity,li2020comment} 

In general, the flexible outcome modeling represents a simple and powerful approach for causal effect estimation, with the caveat that accurate estimation critically depends on correct model specification of the outcome surface. \cite{antonelli2019discussion} With the rapid advancement of machine learning techniques, more precise modeling of the outcome surface can be achieved, which can serve as the building block for improved individualized causal effect estimation. The ISTE estimates can also be utilized to explore the degree of TEH. For example, as in the second stage of the virtual twins approach, the ISTE estimates were used as the outcome and a tree diagram was fitted to identify predictors that explain away differences in the ISTE estimates. In Lu et al.,\cite{lu2018estimating} the ISTE estimates were used as the dependent variable in a standard regression model to explain between-subgroup differences in treatment effect. As suggested in Anoke et al.,\cite{anoke2019approaches} averaging ISTE estimates within a pre-specified subgroup (e.g., female vs. male) can also facilitate exploration of the TEH. Given these existing tools, the ISTE from a flexible outcome model can yield insights not only for the population average treatment effect but also for the detection of TEH. Another interesting practical point regards how to use the estimated ISTE to inform treatment decisions for future patients when only part of the covariates $X^{\text{obs}}$ will be available at the decision time, and $X^{\text{mis}}$ will not be collected. In such situations, a plausible strategy is to marginalize the product of the ISTE functional $f(\omega \mid x)$ and the conditional distribution $f(x^{\text{mis}} \mid x^{\text{obs}})$, over the probability distribution of $X^\text{mis}$.  However, the estimation of the conditional $f(x^{\text{mis}} \mid x^{\text{obs}})$ can be  computationally challenging, especially when there are a large number of covariates.

In the NCDB case study, we showed how the counterfactual ISTE estimates from machine learning models could be utilized to understand whether there might be TEH among localized high-risk prostate cancer patients who were treated with two different radiotherapy approaches. These two radiotherapy techniques did not show significantly different average treatment effects in the overall study population, but we found age and PSA are two key patient factors that modulated the comparative treatment effect. We further exploited the posterior MCMC samples of the ISTE from AFT-BART-NP and the binary tree model CART to explore interesting connections between radiotherapy techniques, patient characteristics and median counterfactual survival. This causal analysis revealed that patients with very high PSA or older patients with low PSA may have enhanced treatment benefit from EBRT+brachy$\pm$AD or EBRT+AD. The results could facilitate treatment effect discovery in subpopulations and may inform personalized treatment strategy and the planning of future confirmatory randomized trials. 

While we have defined our individual-level causal estimand based on the median counterfactual survival times in Section \ref{sec:methods}, it is straightforward to extend this definition to alternative individual-level causal estimands. For example, we can take different values for $q\in(0,1)$ in equation \eqref{eq:ISTE} than $1/2$ and obtain the survival causal effect conditional on the individual profile $X=x$ for other quantiles of the distribution of failure times. While the median survival times are frequently used for its simple interpretation in clinical practice, studying individual quantile survival causal effects could provide additional insights on how treatment effect varies across the distribution of failure times. However, when $q$ departures from $1/2$, the effective sample size for estimating the individual causal effect may decrease due to data sparsity, and the resulting interval estimates could become wider. Beyond the quantile survival causal effect, we can also define individual causal estimands by contrasting the conditional survival probability up to a fixed time $t$, conditional restricted mean survival times (RMST), and conditional (cumulative) hazard rate, extending the population-level estimands introduced in, for example, Chen and Tsiatis \cite{chen2001causal}, Zhang et al.\cite{zhang2012contrasting,zhang2012double}, Bai et al. \cite{bai2013doubly} and Mao et al. \cite{mao2018propensity}. Because we embedded the machine learning approaches considered in the outcome modeling framework and were able to estimate the individual counterfactual survival curves, it is convenient to leverage the estimated individual-specific survival curves and obtain TEH represented by these alternative definitions. We have not examined the corresponding empirical results, and therefore recommend future investigations to identify the best approaches for estimating these alternative individual survival causal effect estimands. 

We acknowledge that there are several limitations of our study, which could stimulate future research in the field of (frequentist or Bayesian) causal machine learning. 
First, while our simulations aimed at providing new empirical evidence about the operating characteristics of state-of-the-art machine learning techniques for estimating TEH with survival data, we only provided frequentist coverage probability for AFT-BART-NP because the credible intervals are readily available from the MCMC output. On the other hand, 
it may be challenging to precisely estimate the variance of the ISTE using DL, RSF and TSHEE, without resorting to more computationally intensive sample-splitting methods. \cite{wager2018estimation,cui2020estimating,diaz2020machine,chernozhukov2018double} Developing and investigating new methods for the variance and interval estimation for frequentist machine learning represent an important avenue for future research. Second, an anonymous reviewer pointed out that one may further improve the estimation of ISTE by parameterizing the counterfactual AFT mean model as $E[\log(T(z))|X=x]=\alpha(x)+(z-e(x))\tau(x),$
where the ISTE can be explicitly represented as a function of $\tau(x)$. This apparent parameterization orthogonalizes the model into two compinents: a prognostic score, $\alpha(x)-e(x)\tau(x)$, as well as a treatment effect model, $z\tau(x)$, and anchors the propensity score in model formulation to build in robustness. While this idea represents a promising approach, developing a formal estimation procedure by placing separate regularization priors on these two model components along the lines of Hahn et al. \citep{hahn2020bayesian} is beyond the scope of this work, and will be pursued in future work. Third, we have assumed conditional noninformative censoring so that the counterfactual censoring time is independent of the counterfactual survival time given baseline covariates. It would be interesting to further develop these machine learning to address informative censoring such that the censoring process is a function of some time-varying covariates.\citep{siannis2005sensitivity} Combining the inverse probability of censoring weights \cite{robins2000correcting,zhang2011estimating} and the flexible outcome models also represents a promising approach to obtain valid causal estimates in this scenario but requires further study in the context of TEH. Finally, as in all causal inference work with observational data, we require an untestable assumption that the observed pre-treatment covariates are sufficient to de-confound the relationship between treatment and outcome. We conducted an \emph{exploratory} sensitivity analysis to demonstrate that, the violation of this assumption will lead to more biased ISTE estimates even for machine learning methods. To wit, using flexible modeling techniques alone will not address violations of structural assumptions (such as weak unconfoundedness) required for causal inference. Developing a formal sensitivity analysis approach to capture the sensitivity of an ISTE estimation method to the potential magnitude of departure from the unconfoundedness assumption would be a worthwhile and important contribution.\citep{hogan2014bayesian}

\section*{ACKNOWLEDGEMENTS}
This work was supported in part by the National Cancer Institute under grant NIH NCI R21CA245855 and grant NIH NCI P30CA196521-01, and by award ME\_2017C3\_9041 from the Patient-Centered Outcomes Research Institute (PCORI). The content is solely the responsibility of the authors and does not necessarily represent the official views of the National Institutes of Health or PCORI. The authors thank the Associate Editor and two anonymous referees, whose suggestions have substantially improved the exposition of this work. 

\section*{DATA AVAILABILITY STATEMENT}
The simulation codes that generate the data supporting the findings of the simulation study are openly available on GitHub at \url{https://github.com/liangyuanhu/ML-DL-TEH}. 
The NCDB data used in the case study is publicly available upon approval of the NCDB Participant User File application. 
\bibliography{TEH}
\end{document}